\documentclass{emulateapj}
\usepackage{lscape}

\newcommand {\apgt} {\ {\raise-.5ex\hbox{$\buildrel>\over\sim$}}\ }
\newcommand {\aplt} {\ {\raise-.5ex\hbox{$\buildrel<\over\sim$}}\ }

\newcommand{\mbh}{M_{\rm BH}}%
\newcommand{\etal}{et al.~}%

\def\sun{\hbox{$\odot$}}
\def\cm3{cm$^{-3}$}

\def\msun{M_{\odot}}
\def\mbh{M_{{\rm BH}}}

\def\beq{\begin{equation}}
\def\eeq{\end{equation}}

\def\sles{\lower2pt\hbox{$\buildrel {\scriptstyle <}
   \over {\scriptstyle\sim}$}}
\def\sgreat{\lower2pt\hbox{$\buildrel {\scriptstyle >}
   \over {\scriptstyle\sim}$}}

\slugcomment{Accepted for Publication in ApJ}
\shorttitle{Flares from Quiescent Supermassive Black Holes}
\shortauthors{Gezari et al.}

\begin{document}

\title{Luminous Thermal Flares from \\Quiescent Supermassive Black Holes}

\author{Suvi Gezari,\altaffilmark{1,8}
Tim Heckman,\altaffilmark{1}
S.~Bradley Cenko,\altaffilmark{2}
Michael Eracleous,\altaffilmark{3}
Karl Forster,\altaffilmark{4}
Thiago S.~Gon\c{c}alves,\altaffilmark{4}
D.~Chris Martin,\altaffilmark{4}
Patrick Morrissey,\altaffilmark{4}
Susan G.~Neff,\altaffilmark{5}
Mark Seibert,\altaffilmark{6}
David Schiminovich,\altaffilmark{7}
and
Ted K.~Wyder\altaffilmark{4}
}

\altaffiltext{1}{Department of Physics and Astronomy,
        Johns Hopkins University,
        3400 North Charles Street,
        Baltimore, MD 21218 \email{suvi@pha.jhu.edu}.}

\altaffiltext{2}{Department of Astronomy, 601 Campbell Hall, University of California, Berkeley, CA  94720-3411.}

\altaffiltext{3}{Department of Astronomy and Astrophysics, The Pennsylvania State University, 525 Davey Lab, University Park, PA  16803.}

\altaffiltext{4}{California Institute of Technology,
        MC 405-47,
        1200 East California Boulevard,
        Pasadena, CA  91125.}

\altaffiltext{5}{Laboratory for Astronomy and Solar Physics,
     NASA Goddard Space Flight Center,
       Greenbelt, MD  20771.}

\altaffiltext{6}{Observatories of the Carnegie Institute of Washington, Pasadena, CA  90095.}

\altaffiltext{7}{Department of Astronomy,
     Columbia University,
         New York, NY  10027.}

\altaffiltext{8}{Hubble Fellow.}


\begin{abstract}
A dormant supermassive black hole lurking in the center of a galaxy will be revealed when a star passes close enough to be torn apart by tidal forces, and a flare of electromagnetic radiation is emitted when the bound fraction of the stellar debris falls back onto the black hole and is accreted.  Although the tidal disruption of a star is a rare event in a galaxy, $\approx 10^{-4}$ yr$^{-1}$, observational candidates have emerged in all-sky X-ray and deep ultraviolet (UV) surveys in the form of luminous UV/X-ray flares from otherwise quiescent galaxies.  Here we present the third candidate tidal disruption event discovered in the \textsl{Galaxy Evolution Explorer (GALEX)} Deep Imaging Survey: a $1.6 \times 10^{43}$ erg s$^{-1}$ UV/optical flare from a star-forming galaxy at $z=0.1855$.  The UV/optical spectral energy distribution (SED) during the peak of the flare measured by \textsl{GALEX} and Palomar Large Field Camera imaging can be modeled as a single temperature blackbody with $T_{bb}=1.7 \times 10^{5}$ K and a bolometric luminosity of $3 \times 10^{45}$ erg s$^{-1}$, assuming an internal extinction with $E(B-V)_{gas}=0.3$. The \textsl{Chandra} upper limit on the X-ray luminosity during the peak of the flare, $L_{X}(2-10$ keV)$< 10^{41}$ erg s$^{-1}$, is 2 orders of magnitude fainter than expected from the ratios of UV to X-ray flux density observed in active galaxies.  We compare the light curves and broadband properties of all three tidal disruption candidates discovered by \textsl{GALEX}, and find that (1) the light curves are well fitted by the power-law decline expected for the fallback of debris from a tidally disrupted solar-type star, and (2) the UV/optical SEDs can be attributed to thermal emission from an envelope of debris located at roughly 10 times the tidal disruption radius of a $\approx 10^{7} \msun$ central black hole.  We use the observed peak absolute optical magnitudes of the flares ($-17.5 > M_{g} > -18.9$) to predict the detection capabilities of upcoming optical synoptic surveys.  
\end{abstract}


\section{Introduction} \label{intro}

Stellar dynamical models predict that once every $10^{3}-10^{5}$ yr a star in a galaxy will pass within the tidal disruption radius of the central black hole, $R_{p}<R_{T}=R_{\star}(\mbh/M_{\star})^{1/3}$, and will be torn apart by tidal forces (Magorrian \& Tremaine 1999; Wang \& Merritt 2004).  After the star is disrupted, at least half of the debris is ejected from the system, and some fraction $(0.1-0.5)M_{\star}$, remains bound to the black hole and is accreted  (Rees 1988; Ayal \etal 2000).  The fallback of debris onto the black hole, and stream--stream collisions between precessing debris orbits (Rees 1988; Cannizzo \etal 1990; Kim \etal 1999) produce a luminous electromagnetic flare that is expected to peak in the ultraviolet (UV)/X-rays (Ulmer 1999).  

Detections of tidal disruption events are of interest because they are a signpost for the presence of a dormant supermassive black hole in the nucleus of a galaxy that is otherwise quiescent and undetectable (Frank \& Rees 1976; Lidskii \& Ozernoi 1979; Rees 1988).  The observed properties of tidal disruption flares have the potential to probe the mass and spin of supermassive black holes in distant galaxies, independently of locally established scaling relations between the mass of the central black hole and its global host galaxy properties (Gebhardt \etal 2000; Ferrarese \& Merritt 2000; Graham \etal 2001; Marconi \& Hunt 2003). Tidal disruption events may also reveal the presence of intermediate mass black holes in globular clusters and dwarf galaxies (Ramirez-Ruiz \& Rosswog 2008).  However, for the most massive black holes, the tidal disruption radius is smaller than the event horizon and stars are swallowed whole without disruption (for a solar-type star $M_{crit} \apgt 10^{8} \msun$, Hills 1975).  Future wide-field time domain surveys that will coincide with the \textsl{LISA} mission have the exciting potential to use the electromagnetic detection of a tidal disruption event to localize the burst gravitational waves that is produced when the star approaches the tidal disruption radius (Kobayashi \etal 2004), and to look for tidal disruption events from recoiling supermassive black holes after a binary merger (Komossa \& Merritt 2008).  

The most convincing candidates for a tidal disruption event are luminous, UV/X-ray flares from galaxies which have no ongoing active galactic nucleus (AGN) activity for which a flare could be attributed to the AGN's central engine.   Strong candidates have now emerged in all-sky X-ray surveys (\textsl{ROSAT} All-Sky Survey (RASS), Donley \etal 2002; {\textsl XMM-Newton} Slew Survey, Esquej \etal 2006) and in the deep UV \textsl{Galaxy Evolution Explorer (GALEX)} Deep Imaging Survey (DIS; Gezari \etal 2006, 2008).  In a systematic search of 2.9 deg$^{2}$ of \textsl{GALEX} DIS FUV and NUV data discovered two luminous UV flares from non active, early-type galaxies (D1-9 at $z=0.326$ and D3-13 at $z=0.3698$) that were also detected in the soft X-rays in follow-up \textsl{Chandra} imaging.  Because the DIS fields overlap with the optical CFHT Legacy Survey (CFHTLS) Deep Survey, we were able to extract simultaneous optical difference imaging light curves of the flares in the $g$, $r$, and $i$ bands with an excellent cadence of days.  The broadband spectral energy distributions (SEDs) of the flares, and their light curves were in excellent agreement with the theoretical expectations for emission from a tidally disrupted main-sequence star onto a central black hole of few times $10^{7} \msun$ (Gezari \etal 2008). 

The new candidate presented in this paper was discovered in five years of \textsl{GALEX} FUV and NUV DIS observations of the 1.2 deg$^{2}$ DEEP2 23h field.  A large amplitude, $\Delta m \sim 2$ mag, FUV flare was detected in 2007 from a source that was not variable in the three yearly epochs prior.  The source, hereafter referred to as D23H-1, was followed up with Keck LRIS optical spectroscopy, which identified the host as a star-forming galaxy.  We identified D23H-1 as a candidate tidal disruption event, and followed it up with optical and X-ray imaging observations nearly simultaneously with the peak of the flare, and months later, and with a late-time optical spectrum one year later. The paper is organized as follows, in Section \ref{obs} we describe the optical, UV, and X-ray imaging and optical spectroscopy of the flare and its host galaxy, in Section \ref{interp} we classify the host galaxy as a star-forming galaxy, and compare the SED and light curve of the flare with what is expected for a tidal disruption event, in Section \ref{demo} we compare the light curves and broadband properties of all three \textsl{GALEX} candidates and compare them to all of the candidates reported from UV and X-ray surveys to date, and in Section \ref{conc} we conclude by using the observed optical properties of the flares to predict the detection capabilities of the next generation of optical synoptic surveys.  Throughout this paper, calculations are made using $H_{0}=70$ km s$^{-1}$ Mpc$^{-1}$, $\Omega_{M}=0.3$, $\Omega_{\Lambda}$=0.7 and a luminosity distance for D23H-1 of $d_{L}=900$ Mpc.  UT dates are used throughout, and magnitudes reported are in the AB system and corrected for a Galactic extinction of $E(B-V)=0.035$.

\section{Observations} \label{obs}
\subsection{\textsl{GALEX} UV Photometry} \label{uv}
\begin{figure*}
\plotone{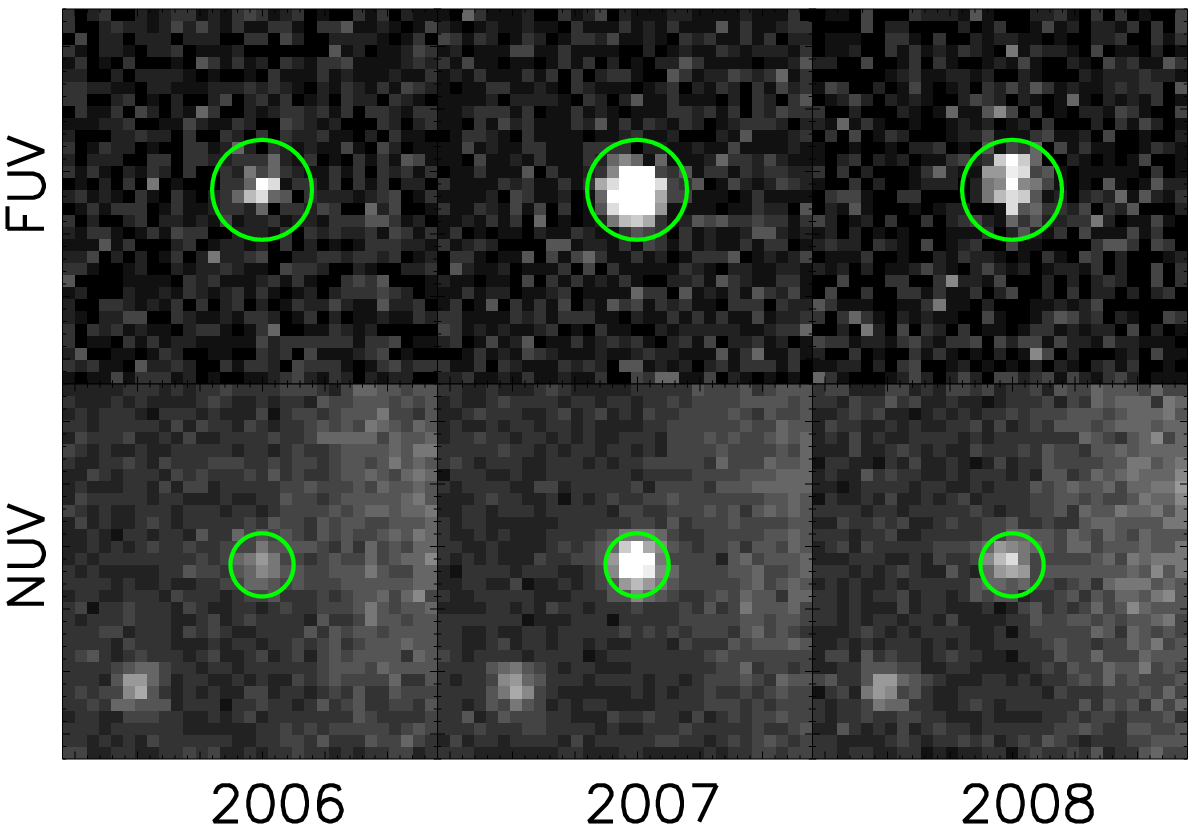}
\caption{\textsl{GALEX} yearly epochs of observations that show the large amplitude flare with a factor of 9.0 increase in flux in the FUV and factor of 4.7 increase in flux in the NUV between 2006 and 2007. Circles show the circular aperture with $r=6\farcs0$ in the FUV and $r=3\farcs8$ in the NUV.
\label{fig:im}
}
\end{figure*}

The \textsl{GALEX} FUV  ($\lambda_{eff}=1539$\AA) and NUV  ($\lambda_{eff}=2316$\AA) observations of the DEEP23H\_02 DIS field from 2004 to 2008 were coadded into nine epochs with exposure times ranging from 6 to 17 ks.  We performed aperture photometry with the center fixed to the centroid of the source in the epoch with the highest signal-to-noise detection, $\alpha = $23h 31m 59$\fs$53, $\delta = +00^{\circ} 17\arcmin 14\farcs$57 (J2000.0).  We used a 6$\farcs$0 and 3$\farcs$8 radius aperture in the FUV and NUV, respectively, with aperture corrections from Morrissey \etal (2007) of 0.87 and 0.58, respectively.  We chose a smaller aperture in the NUV in order to avoid contamination from a bright star located 45$\farcs$8 to the west of the source, and we added 0.19 mag of systematic error in quadrature to the photometric error to account for variable contamination in the aperture by the bright star.    The sky background is assumed to be the pipeline-generated SExtractor catalog sky value at the location of the source (Bertin \& Arnouts 1996) in a reference image constructed from a very deep coadd ($t_{\exp}=100$ ks).  This avoids an overestimation of the sky background that has been found to occur in shallow observations in comparison to deep coadded images (Morrissey \etal 2007).  We determine the photometric errors by measuring the dispersion of the differences of $\sim 5000$ sources in each epoch from the ''true'' magnitude in the reference image as a function of magnitude.  We also determine the mean offset between each epoch and the reference image by measuring the mean of the Gaussian distribution of the difference magnitudes for all sources brighter than 21.5 mag, resulting in offsets for each epoch from the reference image in the range of $-0.039$ to $+0.058$ mag in the FUV $-0.023$ to $+0.020$ mag in the NUV.  Figure \ref{fig:im} shows the FUV and NUV \textsl{GALEX} images during the large amplitude flare, and Table \ref{tab:uv} lists the photometry for each epoch. 

\begin{figure*}
\plotone{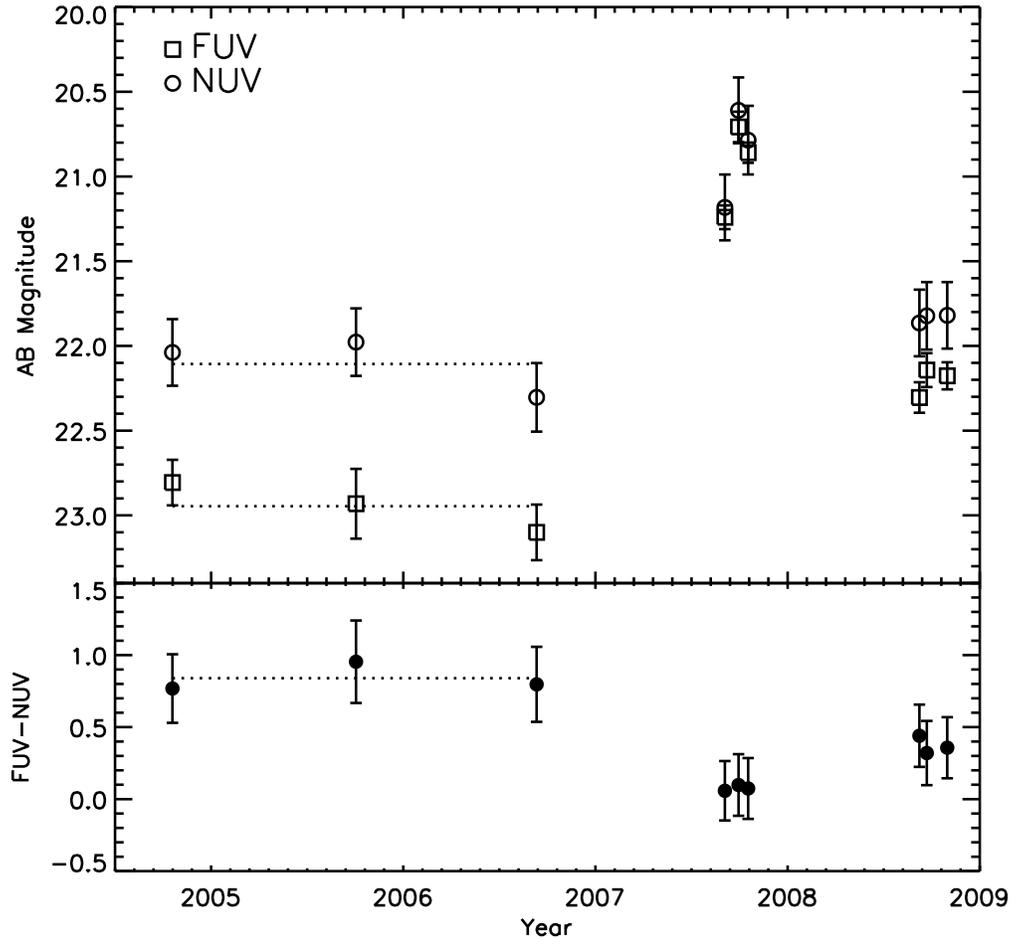}
\caption{{\it Left}: FUV and NUV \textsl{GALEX} light curve, and FUV--NUV color as a function of time, in AB magnitudes corrected for Galactic extinction.  The source is bluer during the ``flaring state'', indicating that the flare has a different SED from the ``steady state''. 
 \label{fig_lc}
}
\end{figure*}

Figure \ref{fig_lc} shows the FUV and NUV light curve of D23H-1 and its FUV$-$NUV color over time.  The flux of the source is constant within 1$\sigma$ in the yearly epochs from $2004-2006$, and the maximum peak-to-peak amplitude of the flare from its steady state on 2006 September 10 to its peak on 2007 September 29 is an increase in flux by a factor of $9.0\pm1.5$ in the FUV and $4.7\pm1.2$ in the NUV.  The FUV$-$NUV color is 0.7 mag bluer during the peak of the flare, indicating that the SED of the source during its ``flaring state'' is different than during its ``steady state''.  The \textsl{GALEX} observations constrain the peak of the flare, $t_{0}$, to be $t_{0}=2007.74 \pm 0.07$.

\subsection{Optical Photometry} \label{opt}
\begin{figure*}
\plotone{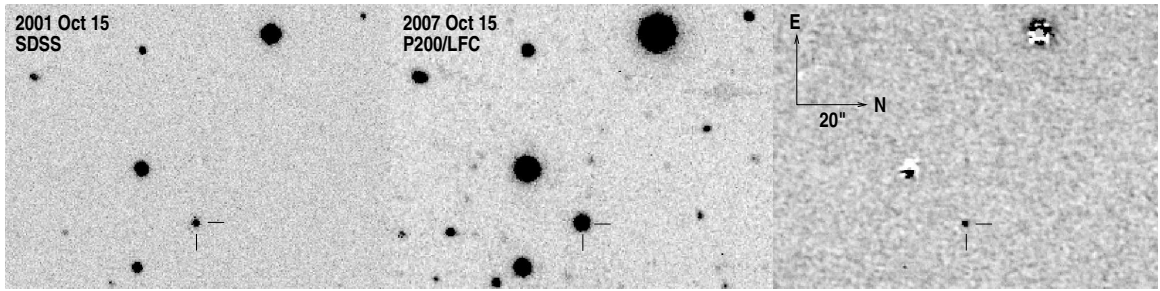}
\caption{Optical imaging of the field of D23H-1.  \textit{Left:} SDSS $g$-band reference image, taken on 2001 October 15 UT.  D23H-1 is indicated with the black tick marks. \textit{Center:} LFC $g$-band image of the same field on the night of 2007 October 15.  \textit{Right:} resulting subtraction, showing a residual point source at the location of the centroid of D23H-1.  The two poorly subtracted stars were saturated in the LFC image.  All images are oriented with north to the right and east up, as indicated by the compass.
 \label{fig:lfc}
}
\end{figure*}

The optical counterpart to the flaring UV source in the DEEP2 DR1 survey catalog from 1999 to 2000 (Coil \etal 2004) is a resolved galaxy with a half-light radius in the $R$ band of 0$\farcs$918 detected with $B = 20.665 \pm 0.005$, $R = 19.296 \pm 0.001$, and $I=18.825 \pm 0.001$, and in the Sloan Digital Sky Survey (SDSS) DR6 catalog from 2001 September with $u=21.52 \pm 0.25, g=20.172 \pm 0.027, r=19.281 \pm 0.018, i=18.859 \pm 0.019$, and $z=18.677 \pm 0.062$ (Adelman-McCarthy \etal 2008). 
We obtained optical imaging in the $g$ band with the Palomar Large Field Camera (LFC) on UT 2007 October 15 (with 1$\farcs$5 seeing) and November 14 (with 1$\farcs$9 seeing) corresponding to 16 and 46 days after the peak of the flare (or $t_{0}+16$ days and $t_{0}+46$ days), with an exposure time of 1620 s and 1800 s, respectively.  The spatial resolution of the ground-based imaging is insufficient for determining the morphology of the galaxy beyond the fact that it is resolved.  Using the SDSS $g$-band images as a template we performed digital image subtraction using the High Order Transformation of PSF And Template Subtraction (HOTPANTS\footnote{http://www.astro.washington.edu/users/becker/hotpants.html}), which is based on the original algorithm of \citet{a00}.   In both epochs we find a residual point source consistent with the location of the host galaxy D23H-1 (Figure \ref{fig:lfc}).  Using several nearby SDSS stars for reference, we measure $g$-band magnitudes of $22.25 \pm 0.12$ (2007 October 15) and $22.37 \pm 0.17$ (2007 November 14) for the flaring source.

\subsection{Spectroscopic Observations} \label{spec}
We obtained an optical spectrum with Keck I LRIS and the 400/8500 grating on 2007 September 16 with a 1$\arcsec$ slit resulting in a spectral resolution of $\Delta\lambda = 5.7$\AA, and a wavelength coverage of $5700-8300$\AA~at $t_{0}-13$ days, and with Keck II DEIMOS and the 1200 line/mm grating on 2007 October 7 with a 1$\arcsec$ slit resulting in $\Delta\lambda = 1.8$\AA~and a wavelength coverage of $4600-7300$\AA, at $t_{0}+8$ days, both with $t_{exp}=1800$ s.  The spectra are shown in Figure 4.  The LRIS spectrum was reduced with standard IRAF routines, and the DEIMOS spectrum was reduced using the DEEP2 DEIMOS data pipeline, which is optimized for data taken with the 1200 line/mm grating as used in this work.  We calibrated the flux scale of the LRIS spectrum by integrating under the SDSS $i$-band filter curve and scaling the flux to match the $i$ magnitude.  We measure the redshift of the host galaxy from the narrow emission lines to be $z=0.1855$. 

The narrow emission-line ratios and their 1$\sigma$ errors determined from Gaussian fits (shown in Figure \ref{fig_spec_zoom}) to the spectra after subtracting a template for the stellar absorption lines, are: [O~III] $\lambda5007$/H$\beta = 0.6 \pm 0.1$, [N~II] $\lambda$6583/H$\alpha = 0.41 \pm 0.03$, ([S~II] $\lambda6716 + \lambda6731$)/H$\alpha=0.22 \pm 0.06$, and an upper limit of [O~I] $\lambda$6300/H$\alpha < 0.035$, using a 3$\sigma$ upper limit on the [O~I] flux measured from propogating the standard deviation of the continuum over 5 pixels.  Figure \ref{fig:bpt} shows the line ratios in the standard diagnostic diagrams (Baldwin, Phillips, \& Terlevich 1981; Veilleux \& Osterbrock 1987, Ho \etal 1997) and in comparison to the empirical dividing line between star-forming galaxies and Type 2 AGNs measured from SDSS (Kauffmann \etal 2003) and the theoretical boundaries between star-forming galaxies, Seyferts and low-ionization emission-line regions (LINERs) from Kewley \etal (2001).  The line ratios fall within the boundaries
for HII regions of star-forming galaxies in all three diagrams.  
In the [O~III]/H$\beta$ versus [N~II]/H$\alpha$ diagram, however,
the object lies very close to the boundary between star-forming galaxies and composite galaxies
from Kewley \etal (2001), and the boundary between star-forming galaxies and AGNs from Kauffmann \etal (2003).  This could allow for some contamination to a pure star-forming spectrum by
an AGN nucleus.  To investigate this, we plot a ``mixing line'' of a pure star-forming
spectrum with a coordinate of ($+0.45, +0.37$) with a LINER nucleus with a coordinate of 
($0.0, +0.3$) with gray dots in Figure \ref{fig:bpt}.  The position of D23H-1 is consistent
with a mixture in which 90\% of the Balmer emission is from a star-forming galaxy, and 10\% 
of the Balmer emission is from a LINER nucleus.  A mixture with a Seyfert 
spectrum at ($+0.1$,$+0.6$) would allow for only 5\% of the Balmer emission to be from a
Seyfert nucleus.
In both cases, the LINER or Seyfert nucleus would contribute $\sim 30$\% to the 
total [O~III] line emission.  Thus, although there may be some contribution from an AGN, 
the narrow emission lines are dominated by star formation.

The spatial profile of the spectrum at the location of H$\alpha$ and at the continuum near a rest wavelength of 6500 \AA~(which is dominated by host galaxy starlight) both have full width at half-maximums (FWHMs) that are equivalent to the seeing during the observation (estimated to be 0$\farcs8-0\farcs$9 from the FWHM in the spatial direction of the standard stars observed before and after the target).  Thus, we do not have the spatial resolution necessary to determine if the H$\alpha$ line is extended and following the spatial distribution of the stellar continuum light (as would be the case if it was tracing star formation), or if it is more compact and coming only from the nucleus (as would be the case if it was powered by an AGN).
 
\begin{figure*}
\plotone{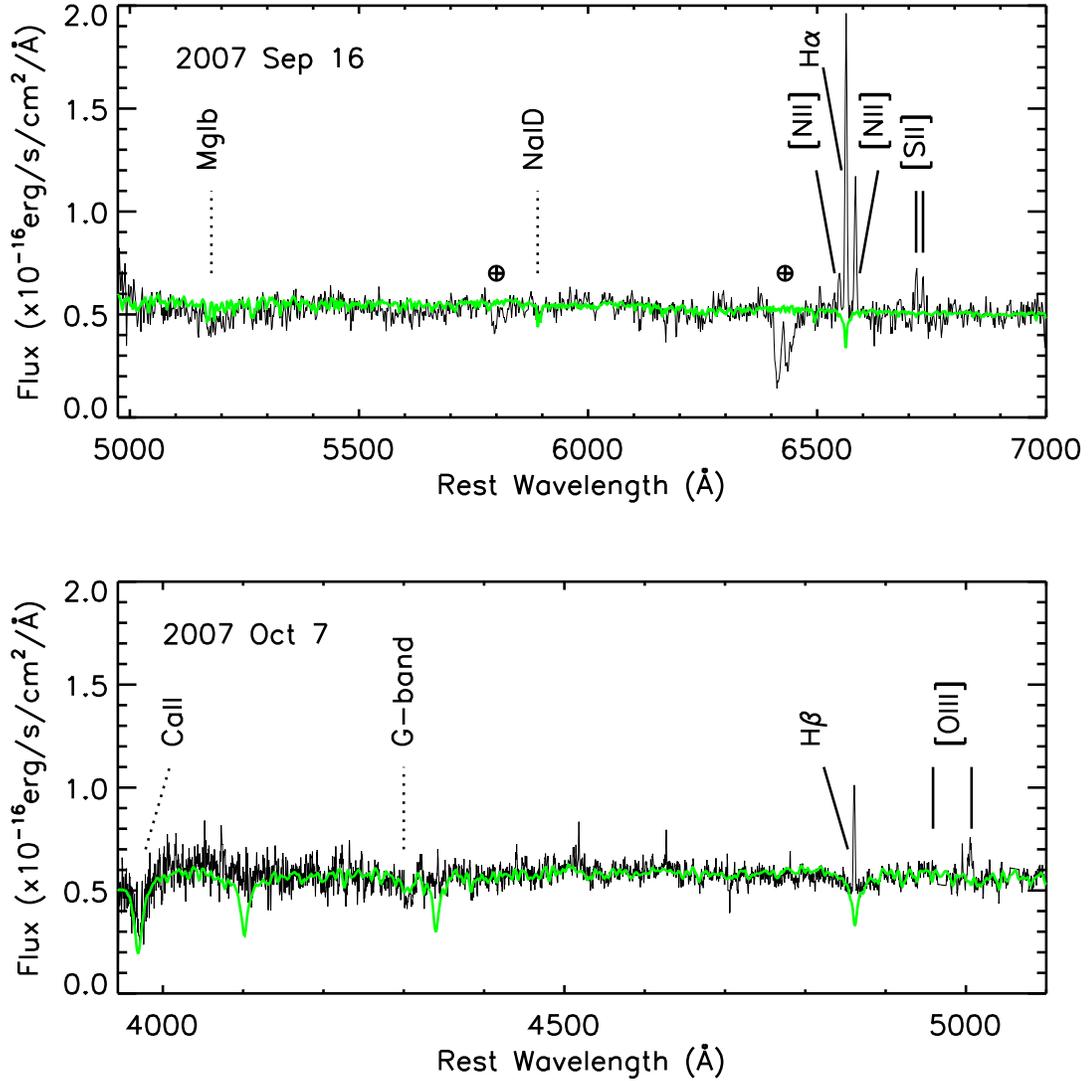}
\caption{
Spectra from LRIS taken 13 days before the peak of the flare, and with DEIMOS 8 days after the peak of the flare.  The best-fitting Bruzual \& Charlot (2003) galaxy template is overplotted in green.  Emission-line features are labeled with the solid tick marks, and stellar absorption features are labeled with the dotted tick marks.  The hatch marks indicate the locations of the strong O$_{2}$ telluric $B$- and $A$-band absorption features.
\label{fig_spec}
}
\end{figure*}

\begin{figure*}
\plotone{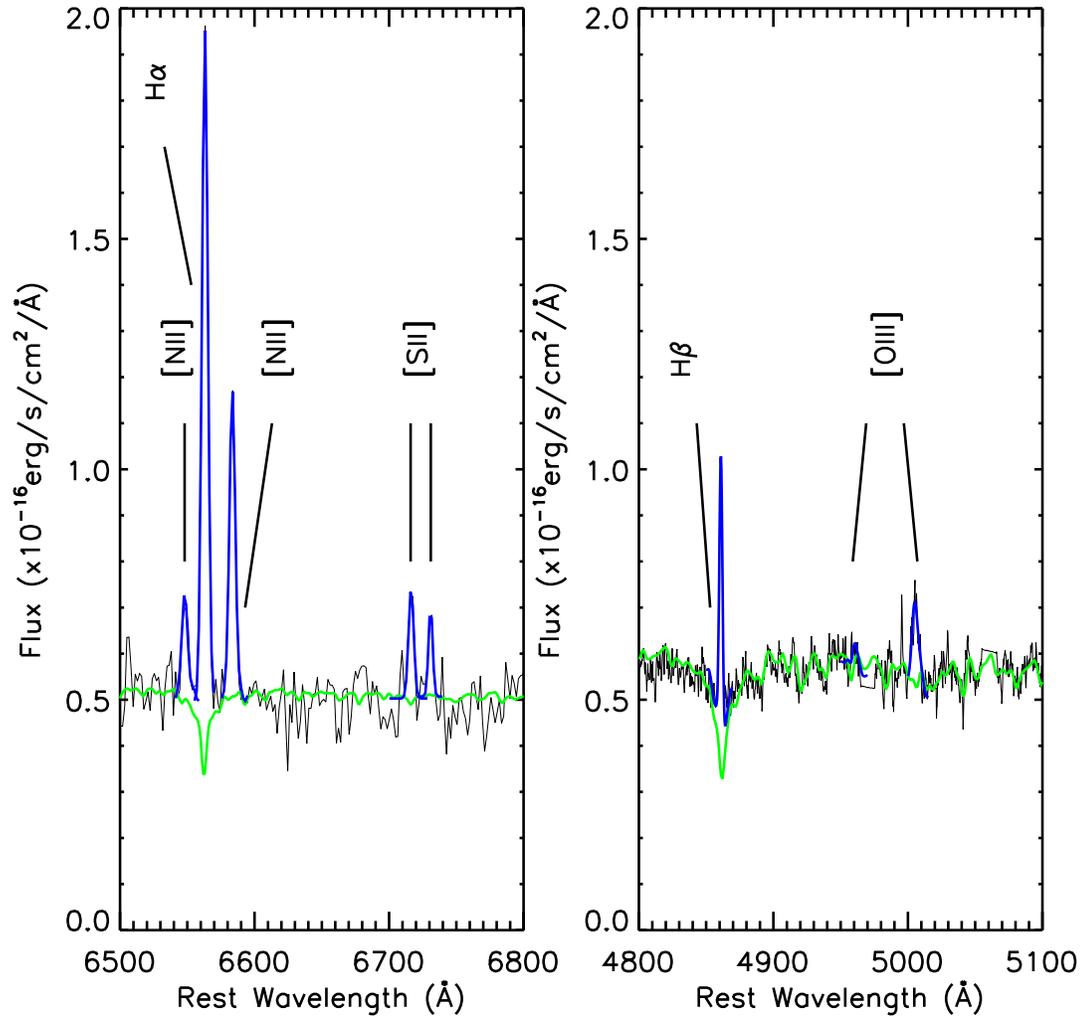}
\caption{
Gaussian fits to the narrow emission lines (shown in blue) after the galaxy template (shown in green) has been subtracted to account for the contribution of stellar absorption lines.
\label{fig_spec_zoom}
}
\end{figure*}

\begin{figure*}
\plotone{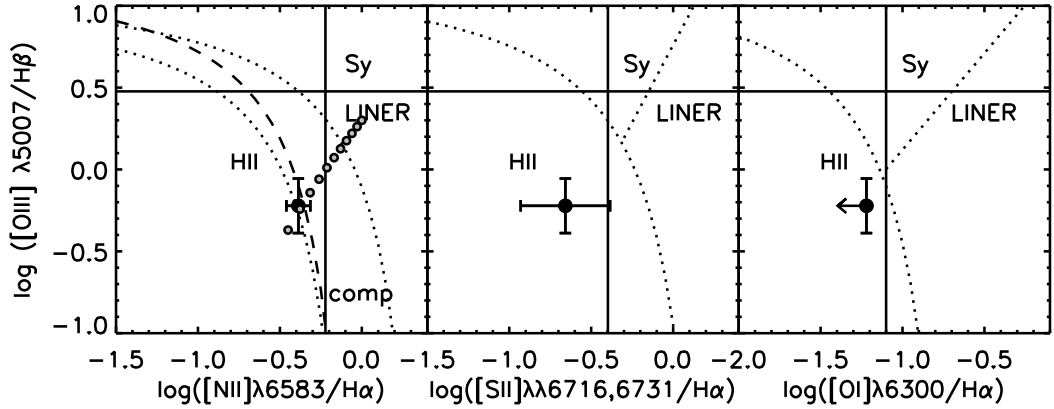}
\caption{
BPT diagram of diagnostic line ratios which classify D23H-1 as a star-forming galaxy.  The empirical dividing line between star-forming galaxies and AGNs from Kauffmann \etal (2003) shown with a dashed line.  Definitions of HII galaxies, LINERs, and Seyfert galaxies from Ho \etal (1997) shown with the solid lines. The dotted lines show the theoretical divisions from Kewley \etal (2001). The region between the dotted lines in the first panel is where composite objects lie. The gray dots show the ``mixing line'' between a pure star-forming spectrum and a LINER spectrum, by adding a contribution from LINER emission to the Balmer emission lines in 10\% increments.  \label{fig:bpt}
}
\end{figure*}

Motivated by the fact that the narrow-line ratios classify the 
galaxy spectrum as a star-forming galaxy, we fit the UV/optical SED measured 
by \textsl{GALEX}, SDSS, and DEEP2 in the 
steady state with a subset of 25 Bruzual \& Charlot (2003 ; BC03) galaxy 
templates with solar metallicity and a single burst of star formation with 
ages from $3.2 \times 10^{5}$ yr to $2.0 \times 10^{9}$ yr.  
For these template fits, we have assumed that the UV and optical flux in
the steady state is dominated by stars in the galaxy.  
We simultaneously fit for internal extinction by adding extinction
to the galaxy templates using the stellar extinction law from Calzetti (2001) 
with $E(B-V)_{stars}=0.44 E(B-V)_{gas}$, for $E(B-V)_{gas} = 0 - 0.6$.  We 
find the least-squares fit for templates which have a specific star-formation rate (SFR) of 
$1.68 \times 10^{-10} \msun$ yr$^{-1}$ $(M_{g}/\msun)^{-1}$, 
where $M_{g}$ is the mass of the galaxy, an age of 2 Gyr, and 
internal extinctions of $0.17 \le E(B-V)_{gas} \le 0.30$.  
Figure \ref{fig_spec} shows the spectra and the best-fitting BC03 galaxy 
template with $E(B-V)_{gas} = 0.3$.  The stellar absorption features 
in the spectrum, including Ca II, $G$-band, Mg Ib, and Na ID, match the 
galaxy template very well, except for the Balmer series lines which are 
filled in from the contribution of narrow-line emission.
Because the DEIMOS spectrum was not corrected for the response of the 
instrument, we divide out the shape of the spectrum, and multiply it 
by the shape of the best-fitting BC03 galaxy template.

Due to the possible contribution of a LINER nucleus at the 10\% level or
a Seyfert nucleus at the 5\% level to the Balmer emission, it is possible
that some fraction of the steady-state
UV flux is non-stellar.  If we repeat the same fits to the steady-state
optical SED, and exclude the \textsl{GALEX} UV photometry,
we can fit for lower extinctions, $0 \le E(B-V)_{gas} \le 0.30$, 
lower specific SFRs of $0.70-1.68\times10^{-10} (M_{g}/\msun)^{-1}$, 
and older ages of $2-3$ Gyr. 
The best-fitting template to the optical data not allowing for
internal extinction is the 
same as the best-fitting template to the UV and optical data allowing for 
internal extinction.  Thus, the UV photometry is entirely consistent with 
the age and SFR fitted to the optical photometry as long as 
there is some internal extinction added.  Therefore, instead
of an {\it excess} of UV flux above the star-forming galaxy template
that would suggest a nonstellar origin, some internal 
extinction in the galaxy is required to explain the observed {\it deficit} in UV flux.

Note that there is an age-metallicity degeneracy, such that higher 
metallicity galaxies
can produce redder optical colors that mimic an older stellar population.
This degeneracy can be broken by comparing the strength of the
metal absorption-line features, which increase in strength with 
higher metallicity (see Figure 10 in BC03).
Given the good match of the solar metallicity BC03 template to the stellar
absorption features in the spectrum of D23H-1, 
invoking an older/lower metallicity or younger/higher metallicity template 
is not necessary.

\subsection{Internal Extinction}

The measured H$\alpha$/H$\beta$ ratio is sensitive to the galaxy template chosen to subract the stellar continuum.  If we use the best-fitted BC03 template to the broadband photometry, and add the range of best-fitted internal extinctions, $0.17 \le E(B-V)_{gas} \le 0.30$, this results in H$\alpha$/H$\beta=4.1-4.4$.  These values are larger than expected for case B recombination ([H$\alpha$/H$\beta]_{0}=2.87$).  Because the [O~III]/H$\beta$ versus [N~II]/H$\alpha$ line ratio for D23H-1 is close to the boundary between star-forming galaxies and AGNs shown in Figure \ref{fig:bpt}, it may be that D23H-1 has a ``composite spectrum'', with some contamination by an AGN.   In AGN spectra 
there is a contribution to the narrow H$\alpha$ line from collisional excitation that increases the Balmer line ratio to (H$\alpha$/H$\beta)_{0}=3.1$ (Osterbrock 1989).  The observed Balmer decrement is $R_{\alpha\beta}$ = (H$\alpha$/H$\beta)_{0}$/(H$\alpha$/H$\beta)_{obs}$ = $1.4-1.5$ when assuming (H$\alpha$/H$\beta)_{0}=2.87$, and $1.3-1.4$ when assuming (H$\alpha$/H$\beta)_{0}=3.1$, and thus (including all the systematic uncertainties) implies $E(B-V)_{\rm gas}=0.3\pm0.1$ using Table 2 in Calzetti (2001).  It is encouraging that both the BC03 template fits to the broadband photometry and the Balmer decrement indicate $E(B-V)_{gas}\sim0.3$.

The observed Galactic extinction-corrected H$\alpha$ luminosity, L(H$\alpha)=8.3\times10^{40}$ erg s$^{-1}$, translates to an unextincted line luminosity of $L_{\rm{H}\alpha}=(1.2 \pm 0.1) \times 10^{41}$ erg s$^{-1}$, and assuming that the H$\alpha$ line is powered by hot stars, an SFR$=(0.95 \pm 0.08) \msun$ yr$^{-1}$ (Kennicutt 1998).  
The $B$-band absolute magnitude of the galaxy is $M_{B}=B-DM-K-A_{B}= -18.4 \pm 0.2$, where we have used the conversion $B=B_{AB}+0.163$, $K=-1.24$ measured from the galaxy template, $DM = 5$log$(d_{L}/10 $pc), and an extinction of $A_{B}=2.22E(B-V)_{gas} = 0.7 \pm 0.2$ using the extinction curve for starlight from Calzetti (2001).  This absolute magnitude corresponds to a galaxy mass from the best-fitting BC03 template of $(4 \pm 1) \times 10^{9} \msun$, and an SFR = $(0.7 \pm 0.2) \msun$ yr$^{-1}$, which is in agreement with the SFR estimated from the unextinguished H$\alpha$ luminosity within the
inherent uncertainties in such techniques (e.g., at the factor of 2 level).

\subsection{Late-time Spectrum}

We obtained a follow-up spectrum 1 yr after the peak of the flare to look
for high ionization He II$\lambda 4686$ or iron lines that could appear as a light echo to a flare in the EUV/soft X-ray photoionizing continuum, as was reported to be the cause for fading high ionization lines in a broad-line AGN by Komossa \etal (2008).
D23H-1 was observed with the Low-Resolution Spectrograph (LRS) and
the G1 grism on the Hobby--Eberly Telescope (HET) on three separate
occasions: 2008 November 3, 17, and 18. On each observation two
1200 s exposures were taken through a $2\arcsec$ slit, oriented
at the parallactic angle, with a final spectrum with $\Delta\lambda = 14~\AA$, 
and a wavelength range of 4090--10,830~\AA.
The reductions were carried out in a standard manner.  Correction of telluric absorption
bands and (relative) flux calibration was carried out using the
spectra of spectrophotometric standard stars observed on the same
night and reduced in the same manner as the spectra of the source.  

The spectrum shows neither the appearance of new emission lines, nor any change in line ratios since the spectra taken in 2007. Since the spectrum is still dominated by the host galaxy light, this suggests that (1) there is not enough gas within a light-travel time delay of 1 yr from the central black hole to reverberate the flaring continuum or (2) that the bulk of the flare emission did not extend into the EUV/X-ray wavelength range (hundreds of eV) that is necessary to excite these high-ionization lines.

\subsection{\textsl{Chandra} X-ray Observations} \label{X-ray}
We obtained 30 ks target of opportunity (TOO) observations with \textsl{Chandra} ACIS-S on 2007 October 1--3 ($t_{0}+3$ days) and 2008 January 23 ($t_{0}+116$ days).  Using a circular aperture with a radius of 4 pixels and a background annulus with radii of 8 and 12 pixels, the galaxy has a $<2\sigma$ detection above the background from $0.2-10$ keV of $(1.3 \pm 0.8) \times 10^{-4}$ cts s$^{-1}$, or a 3$\sigma$ upper limit of $< 2.4 \times 10^{-4}$ cts s$^{-1}$.  This corresponds to an unabsorbed flux density at 1 keV of $< 1.7 \times 10^{-33}$ erg cm$^{-2}$ s$^{-1}$ Hz$^{-1}$, for an AGN power law with $\Gamma = 1.7$, and a Galactic absorption column density of $N_{H}=2.03 \times 10^{20}$ cm$^{-2}$.  The total unabsorbed flux in the $0.2-2$ keV band is $<8.5 \times 10^{-16}$ erg cm$^{-2}$ s$^{-1}$ which corresponds to $L(0.2-2$keV$)< 7.2 \times 10^{40}$ erg s$^{-1}$, and in the $2-10$ keV band is $< 1.1 \times 10^{-15}$ which corresponds to $L(2-10$keV)$ < 9.3 \times 10^{40}$ erg s$^{-1}$.  On 2008 January 23 the galaxy was not detected above the 1$\sigma$ level with a 3$\sigma$ upper limit of $2.1 \times 10^{-4}$ cts s$^{-1}$, corresponding to an upper limit to the flux density at 1 keV of $< 1.4 \times 10^{-33}$ erg cm$^{-2}$ s$^{-1}$ Hz$^{-1}$, and an upper limit to the $0.2-10$keV luminosity of $<1.4 \times 10^{41}$ erg s$^{-1}$.  If we assume that the X-ray emission suffers the same amount of extinction as the gas in the HII regions, then for $N_{H}=(1.7 \pm 0.6) \times 10^{21}$ cm$^{-2}$, the X-ray luminosities get multiplied by a factor of $1.3-1.5$.

\section{Interpretation} \label{interp}

\subsection{Nature of the Host Galaxy} \label{sed}

The correlations between SFR and X-ray luminosity  (Hornschemeier \etal 2005) and between $L_{\rm{H}\alpha}$ and $L_{X}(0.5-2 $keV)  (Zezas 2000; Georgakakis \etal 2006) observed for star-forming galaxies would both imply $L_{X} \sim 1 \times 10^{40}$ erg s$^{-1}$ for D23H-1,
which is consistent with the observed upper limit  to the steady-state X-ray luminosity.  If we were to assume that all of the H$\alpha$ luminosity were instead powered by an AGN, which is not consistent with the diagnostic narrow-line ratios which indicate that the object is at most a composite object, then the mean relation between $L_{H\alpha}$ versus $L_{X}(2-10$keV) for AGNs would predict a hard X-ray luminosity of L($2-10$keV)$\approx 10^{42}$ erg s$^{-1}$, which is a factor of 10 above the observed hard X-ray upper limit (Terashima \etal 2000; Ho \etal 2001).  However, there is a significant scatter in the $L_{X}$ versus $L_{H\alpha}$ relation at lower luminosities ($L_{H\alpha}<10^{41}$ erg s$^{-1}$), and so we cannot rule out that the H$\alpha$ line is being powered by an AGN from $L_{X}/L_{\alpha}$ alone.
Finally, if the total [O~III] luminosity were powered by an AGN, then one would expect a hard X-ray luminosity of $> 10^{42}$ erg s$^{-1}$ (Heckman \etal 2005), which again is not observed.  

The lack of broad emission lines or high excitation narrow-line ratios in the host galaxy spectrum disfavor the presence of a normal AGN.  However, UV variability has been associated with LINER galaxies (Maoz \etal 2005; Totani \etal 2005), and used as evidence that a low-luminosity AGN powers the low-ionization emission lines, as opposed to a young stellar continuum.  The largest amplitude NUV variability in these galaxies reach factors of $3-5$ (NGC 4552, Renzini \etal 1995; NGC 4579, Barth \etal 1996; NGC 1399, O' Connell \etal 2005; NGC 3368, NGC 3998, NGC 4203 Maoz \etal 2005), with luminosities of $\sim 10^{39-41}$ erg s$^{-1}$.  Although the amplitude of NUV variability of these objects
are comparable to that observed for D23H-1, their UV fluxes were measured using high spatial resolution \textsl{Hubble Space Telescope (HST)} imaging that isolates the nucleus from the host galaxy UV light.  Our \textsl{GALEX} images in the NUV have a spatial resolution of $5\arcsec$, and can include UV starlight in the aperture.  Given that we successfully fit the steady-state UV/optical SED with starlight alone, with no contribution from an AGN, it is likely that some if not all of the steady-state UV flux is from starlight.  Thus, the observed variability amplitudes for D23H-1 in the FUV and NUV are merely {\it lower limits} to the intrinsic nuclear variability. 

BL Lacs a rare subclass of AGNs that have weak emission lines, but are characterized by strong non-thermal continuum emission and rapid variability.  Their peculiar properties are explained as the result of the orientation of a relativistic jet in these sources toward the direction of the observer (e.g., Blandford \& Rees 1978).  The difficulty with this scenario as an explanation for the lack of AGN spectral features in D23H-1, is that it is not detected in the FIRST radio catalog (White \etal 1997), with an upper limit to the flux at 1.4 GHz of 0.75 mJy.  The corresponding radio-to-optical spectral index, $\alpha_{ro}$, defined by $\alpha_{\nu_{2}\nu_{1}}=-{\rm log}(f_{\nu_{1}}/f_{\nu_{2}})/{\rm log}(\nu_{1}/\nu_{2})$, where $\nu_{1}=4400$ \AA, and $\nu_{2}=1.4$ GHz, is $\alpha_{ro}$=0.24, and places the galaxy in the radio quiet AGN category, and below the values seen even for BL Lacs with high-energy cutoffs (Beckmann \etal 2003). Furthermore, the characteristic SED of a BL Lac has a parabolic shape that extends from the radio to the X-rays.  The lack of a non-stellar continuum diluting the absorption features in the optical spectrum simultaneous with the peak of the flare is also inconsistent with this picture.  Finally, the hosts of BL Lacs are typically massive elliptical galaxies, and not $\approx 10^{10} \msun$ star-forming galaxies.

The UV/optical-to-X-ray spectral index observed during the flare, defined in the literature with $\nu_{1}=1$ keV or 2 keV and $\nu_{2}=2500$~\AA is $\alpha_{ox} > 2.4\pm0.2$.  Here, we are using the extinction-corrected NUV flux density $f_{\nu_{e}}=10^{A_{\nu}/2.5}f_{\nu_{o}}/(1+z)=(5.6 \pm 2.0) \times10^{-28}$ erg s$^{-1}$ cm$^{-2}$ Hz$^{-1}$, where $A(2316$\AA$/(1+z))=3.95E(B-V)_{gas}=1.2\pm0.4$; and the extinction-corrected X-ray flux density $f_{1 keV(1+z)}< 10^{A_{X}/2.5}f_{1 keV}/(1+z) = (2.0\pm 0.1) \times10^{-33}$ erg s$^{-1}$ cm$^{-2}$ Hz$^{-1}$, where $10^{A_{X}/2.5}=(1.4\pm0.1)$.  Without the internal extinction correction of a factor of $3\pm1$ in the NUV and $1.4\pm0.1$ in the X-rays, $\alpha_{ox}$ is $>2.2$.  The observed upper limit to the X-ray flux density is 2 orders of magnitude fainter than expected for Seyfert 1 galaxies in the same luminosity class, which have $\alpha_{ox} =1.0-1.6$ (Strateva \etal 2005; Steffen \etal 2006) or BL Lacs ($\alpha_{ox} = 0.6 - 1.6$, Plotkin \etal 2008), but in the range expected for normal galaxies ($\alpha_{ox} > 1$).  One can argue that the spectral index observed during a flaring state cannot be compared to the average spectral indices observed in AGNs.  There may be a time delay between fluctuations in the UV band and the X-ray band, or spectral
changes that occur in AGNs during high states of luminosity.  It is a known property
of AGNs that they have ``bluer'' colors when they brighten in the optical (Giveon \etal 1999; Geha \etal 2003).  We also found the same behavior in the FUV$-$NUV colors during the high state of variable AGNs and quasars (Gezari \etal 2008).  The implications of $\alpha_{ox}$ during the
steady state are not clear, since we do not know what fraction (if any) of the steady-state NUV flux is from an AGN.  If, for example, we assume that 10\% of the steady-state NUV flux is nonstellar,
then $\alpha_{ox}$ of the nonstellar component is $> 1.5$, and is in the observed range for AGNs 
with $\nu L_{\nu}(2500$\AA)$ \sim 10^{42}$ erg s$^{-1}$ (see Figure 2 in Maoz 2007).  Thus, the UV and X-ray properties of the source during its steady state do not rule out the presence of a low-luminosity AGN. 

Considering that the host of the flare is a star-forming galaxy, we should also investigate the possibility that the transient source is a supernova.  Hydrogen-rich core-collapse supernovae (Type II SNe) are hot ($T_{eff} \sim 10^{4}$ K) and bright in the UV at early times (i.e. SN 1987A, Hamuy \etal 1988; SN 2006bp: Immler \etal 2007, Dessart \etal 2008; SN 2005cs, Dessart \etal 2008; SN 2008es, Gezari \etal 2009).  The SN ejecta cool through expansion and radiation, causing the peak of the SED to shift to optical wavelengths, and the UV emission to drop rapidly on the timescale of weeks.  Although SNe IIn have a UV excess above the cooling blackbody emission that is powered by interaction with circumstellar material that can last for years (i.e., SN 1993J and SN 1998S, Fransson \etal 2005), this UV emission is dominated by emission-line features in a nebular phase, and is a small fraction of the continuum emission from the ejecta.  Thus, the timescale of the UV flare presented here of over a year, and the small change in the FUV$-$NUV color during that time, is not consistent with the rapid cooling associated with SN ejecta.  These arguments in combination with the fact that the position of the source in the UV during all of the \textsl{GALEX} epochs is constant within the astrometric accuracy of $0\farcs5$ (for a galaxy at $d_{L}=900$ Mpc, $0\farcs5$ corresponds to 2 kpc), and that the UV source and the optical flaring source measured from the difference image are coincident with the centroid of the host galaxy favors the association of the flare with the central supermassive black hole.

\subsection{SED of the Flare} \label{flare}
We now analyze the SED of the flare, by subtracting off the ``steady-state''  FUV, NUV, and $g$ band data points from the ``flaring-state'' flux, and fit the remaining SED with a single temperature blackbody in the rest frame of the galaxy.  Assuming that the steady-state SED is dominated by star formation, we correct the fluxes for an internal extinction of $E(B-V)_{gas}=0.3$ using the Calzetti (2001) law.  Since the flaring continuum is associated with the central black hole, the extinction in the line of sight to the galaxy nucleus may not be the same as the global extinction to the galaxy measured from the Balmer decrement of the HII region emission lines.  Thus, to be conservative, we examine both the extreme case of no extinction in the line of sight to the galaxy nucleus, together with the more probable case of an extinction equal to the global extinction in the galaxy.

\begin{figure*}
\plotone{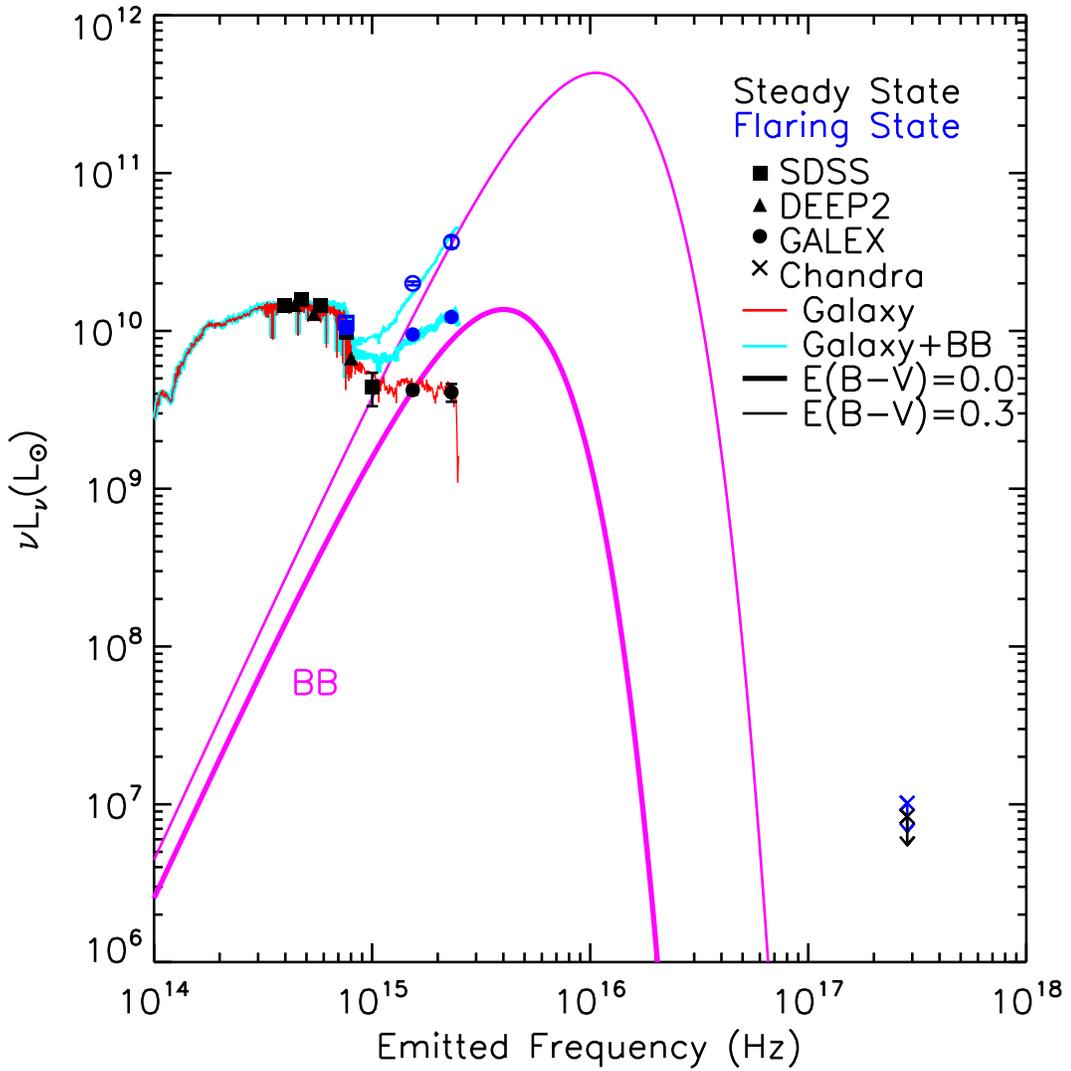}
\caption{SED of the source in the ``steady state'' (black symbols) and in the ``flaring state'' (blue symbols) as a function of frequency shown in the rest-frame of the host galaxy at $z=0.1855$.  The fluxes have been corrected for Galactic extinction of $E(B-V)=0.035$ and extinction internal to the host galaxy of $E(B-V)_{gas}=0.3$ using the Calzetti (2001) law.  The best-fitting Bruzual \& Charlot (2003) galaxy template to the ``steady-state'' SED is shown in red.  A blackbody spectrum  has been added to the host galaxy template to match the ``flaring-state'' FUV, NUV, and $g$-band flux with $T_{bb}=1.7 \times 10^{5}$ K (thin line) when flare fluxes are corrected for internal extinction (open symbols), or $T_{bb}=4.9 \times 10^{4}$ K (thick line) when not corrected for internal extinction (filled symbols).  The \textsl{Chandra} upper limits on the X-ray flux densities at 1 keV are plotted with an X.
 \label{fig_sed}
}
\end{figure*}

The least-squares fit to a blackbody assuming $E(B-V)_{gas}=0.3$ and the Calzetti \etal (2001) law, with internal extinction correction factors of 4.0, 3.0, 2.0 in the FUV, NUV, and $g$ band, respectively, yields $T_{bb}=1.7^{+2.2}_{-0.2} \times 10^{5}$ K and log($L_{bb}$(erg s$^{-1}$))=$45.0^{+1.3}_{0.5}$.  The steady state SED fitted with the BC03 galaxy template, and the blackbody model fit to the flare added to the galaxy template are shown in Figure \ref{fig_sed} in units of $\nu_{e}L_{\nu_{e}}$ as a function of emitted frequency, $\nu_{e}=(1+z)\nu_{o}$, where $L_{\nu_{e}}=f_{\nu_{o}}4\pi d_{L}^2/(1+z)$. Alternatively, the power-law fit to the UV/optical flare SED yields $\alpha = +1.8 \pm 0.4$, where $f_{\nu} \propto \nu^{\alpha}$, and $L(1350-8500$\AA)$ = 5.0\times10^{43}$ erg s$^{-1}$.  Also shown in Figure \ref{fig_sed} is the case for no extinction in the line of sight, which yields $T_{bb}=4.9^{+0.7}_{-0.5}\times10^{4}$ K, log($L_{bb}$(erg s$^{-1}$))=$43.4^{+0.2}_{0.1}$, or $\alpha = +1.2 \pm 0.4$ and $L(1350-8500$\AA)$ = 1.6\times10^{43}$ erg s$^{-1}$.

\subsection{Nature of the UV/Optical Flare}

While there is no {\it direct} evidence for an active nucleus in the host star-forming galaxy of D23H-1, we cannot definitively rule out the presence of a low-luminosity AGN with $L_{X}<10^{41}$ erg s$^{-1}$.  There are two scenarios associated with the host galaxy's central supermassive black hole (SMBH) that could produce the luminous ($>10^{43}$ erg s$^{-1}$) UV/optical flare:  (1) a flaring low-luminosity AGN with peculiar X-ray and optical emission-line properties during its flaring state that differ significantly from measured average properties of AGNs and (2) the tidal disruption of a star by an otherwise quiescent SMBH.  The important distinction between these two scenarios is the presence or lack of a steady-state accretion disk around the central black hole.  

In the case of a tidal disruption event, the UV flare is caused by the sudden increase in the supply of gas on to the central black hole from a stream of tidally disrupted stellar debris that forms a thick accretion disk or torus, and emits a soft, thermal spectrum with $T_{eff} \approx (L_{Edd}/4\pi\sigma R_{T}^{2})^{1/4}$ (Ulmer 1999).   In the accretion disk case, a UV flare is produced by an upward fluctuation in the accretion rate, possibly due to a magnetohydrodynamic (MHD) instability (see review by Schramkowski \& Torkelsson 1996).  The timescale of variability due to accretion disk instabilities has been investigated (Kawaguchi \etal 1998; Hawkins 2002),  and rest-frame NUV variability studies of the ensemble statistics of sparsely sampled optical light curves of tens of thousands of quasars are in general agreement with the slope of the structure functions in these models (Vanden Berk \etal 2004; Bauer \etal 2009).  While a disk spectrum during episodes of higher accretion rates is expected to have a higher effective temperature, and thus appear ``bluer'' (e.g., Shakura \& Sunyaev 1973), there are few theoretical predictions for the corresponding changes in the X-ray properties of the disk.   

There are only a handful of AGNs for which simultaneous UV and X-ray variability has been studied in detail (Uttley \etal 2003; Ar\'evalo \etal 2008; Breedt \etal 2009), and they demonstrate a strong correlation between X-ray and optical flux variations on time-scales of days to months with no measured time-lag.  Reverberation mapping experiments of tens of nearby AGNs, meanwhile, measure a time lag between the UV/optical continuum and the fluxes of broad, recombination lines that reprocess that continuum on the order of weeks to months, corresponding to the light-crossing time of the broad-line region (Peterson \etal 2004).  Future wide-field time domain X-ray surveys, such as EXIST (Grindlay 2004), in coordination with the \textsl{GALEX} Time Domain Survey in the UV, and optical synoptic surveys, such as Large Synoptic Survey Telescope (LSST), could place much needed observational constraints on the relationship between optical, UV, and X-ray variability in a larger sample of AGNs.

Given that both the UV/optical SED and the X-ray-to-UV flux ratio during the flare in D23H-1 are anomalous compared with the average properties of AGNs,  we investigate further the possibility that the flare was a result of the tidal disruption of a star by an otherwise quiescent central black hole, with the caveat that the temporal behavior of these properties are not well constrained both theoretically or observationally in the alternative case of a large amplitude, high luminosity AGN flare.

\subsection{Light Curve} \label{lc}
The light curve of a tidal disruption event should follow the power-law decay of the mass accretion rate of the tidal disruption debris.  The evolution of the mass accretion rate is predicted by analytical calculations (Phinney 1989) and numerical simulations (Evans \& Kochanek 1989; Ayal \etal 2000) to decline as $[(t-t_{D})/(t_{0}-t_{D})]^{-n}$, where $n=5/3$, $t_{0}$ is the time of the peak of the flare, and $t_{D}$ is the time of disruption.  
We perform a least-squares fit of a power-law decay to the FUV light curve of D23H-1, after subtraction of the steady state FUV flux.  In the fit, $t_{0}$ is constrained from the observations, $t_{D}$ is unknown and is a free parameter in the fit, and the power-law index is fixed to $n=5/3$.  This fit yields a time of disruption of $t_{D}=2007.24 \pm 0.06$.  If we instead allow the power-law index to vary, we get $t_{D}=2007.4 \pm 0.4$ and $n=1.3 \pm 0.9$. 

We can place a strict upper limit of $t_{D} < 2007.67$ from the UV detection before the peak of the flare, so we assume $t_{D}=2007.4^{+0.27}_{-0.4}$ yr, which yields $(t_{0}-t_{D})/(1+z) = 0.06 - 0.6$ yr.  This time delay is related to the central black hole mass as $\mbh = k^{3} (R_{p}/R_{T})^{-6} r_{\star}^{-3} m_{\star}^{2} [(t_{0}-t_{D})/0.11 $yr$]^2 \msun$ (Li \etal 2002), where $k$ ranges from 1 to 3 depending on how much the star is spun-up upon breakup, $r_{\star}=R_{\star}/R_{\odot}$, and $m_{\star}=M_{\star}/\msun$, and so we get a range of possible black hole mass solutions that depend of the spin-up and type of the star disrupted, and its penetration factor ($R_{p}/R_{T}$), $\mbh = (0.3-30) \times 10^{6} \msun k^{3} (R_{p}/R_{T})^{-6} r_{\star}^{-3} m_{\star}^{2}$.

In Figure \ref{fig:lc}, we also show a fit to the analytical equation from Lodato \etal (2009) (L09) for the mass accretion rate of a tidally disrupted solar-type star with a polytropic index, $\gamma = 1.4$, which takes into account the internal structure of the star in the calculation.  The instantaneous power-law index of the calculated mass accretion rate is $<1$ at early-times, and asymptotically approaches $n=5/3$ at late times.   We take their equation for $\dot{M}(t)$ for solar-type star disrupted by a $10^{6} \msun$ black hole at $R_{p}=R_{T}$, and scale it by $k^{-3/2}M_{6}^{1/2}(1+z)$, where $M_{6}=\mbh/10^{6} \msun$.  There is a degeneracy in the fit between the mass of the black hole and the spin-up of the star, since decreasing the spin of the star or increasing the mass of the black hole both result in stretching out the light curve in time.  If we assume that the star is spun-up to near break-up upon disruption, $k=3$, then $\mbh = (5.4\pm0.4) \times 10^{7} \msun$, and $t_{D}=2007.56\pm0.01$.  

\subsection{Central Black Hole Mass} \label{mbh}
The black hole mass inferred from the relation between $\mbh$ and $M_{bulge}$ (Magorrian \etal 1998), using the galaxy mass derived from the galaxy template corresponds to $M_{BH} \approx 3 \times 10^{7} (B/T)^{1.13} \msun$, where ($B/T$) is the bulge-to-total luminosity ratio, which is unknown.  For a star-forming galaxy with $\approx 10^{10} \msun$ it is likely a spiral galaxy and $(B/T)$ can range from 0.25 in Sb galaxies to 0.01 for Sd galaxies, or in the case of M33 can be almost a pure disk galaxy ($B/T < 0.01$, van den Bergh 1991).  Spiral galaxies also appear to have systematically lower black hole masses for their given bulge mass compared with the Magorrian \etal (1998) relation (Salucci \etal 2000).  Greene \& Ho (2005) demonstrated that the velocity width of the [O~III]$\lambda$5007 emission line ($\sigma_{\rm [O~III]}$) can be used as a proxy for the velocity dispersion of the stars in the host galaxy bulge ($\sigma_{\star}$).  We find $\sigma_{\rm [O~III]} = 140 \pm 20$ km s$^{-1}$ in the galaxy rest frame, where $\sigma_{\rm [O~III]} = \sqrt{\sigma_{obs}^{2}-\sigma_{inst}^{2}}$.  Using the $\mbh-\sigma_{\star}$ relation from Merritt \& Ferrarese (2001), and including the rms scatter in the log$(\sigma_{\rm [O~III]}) - $log$(\sigma_{\star})$ relation of $\sim 0.2$ (Nelson \& Whittle 1996), this velocity dispersion corresponds to a central black hole mass of log$(\mbh/msun)=(7 \pm 1)$.  While the error bars of this estimate are large, it is encouraging that the black hole mass range is within the masses for which the tidal disruption of a solar-type star is observable.

Assuming that $L(t) = L_{0}[(t-t_{D})/(t_{0}-t_{D})]^{-1.3}$, where we set $L_{0}$ to the lower-bound to the peak blackbody luminosity, $L_{0} = 3 \times 10^{44}$ erg s$^{-1}$, and $t_{0}=2007.74$, and $t_{D}=2007.4$, then the total energy radiated found by integrating the curve from $t_{0}$ out to infinity is $1 \times 10^{52}$ erg, and corresponds to a total material accreted of $M_{acc} = E_{tot}/(\epsilon c^{2}) = 0.06 \msun$, for $\epsilon=0.1$, and $M_{\star} > 0.1 \msun$ if at least half of the stellar debris is expelled after disruption and escapes the system. 

\section{Demographics of Tidal Disruption Event Candidates} \label{demo}

\begin{figure*}
\plottwo{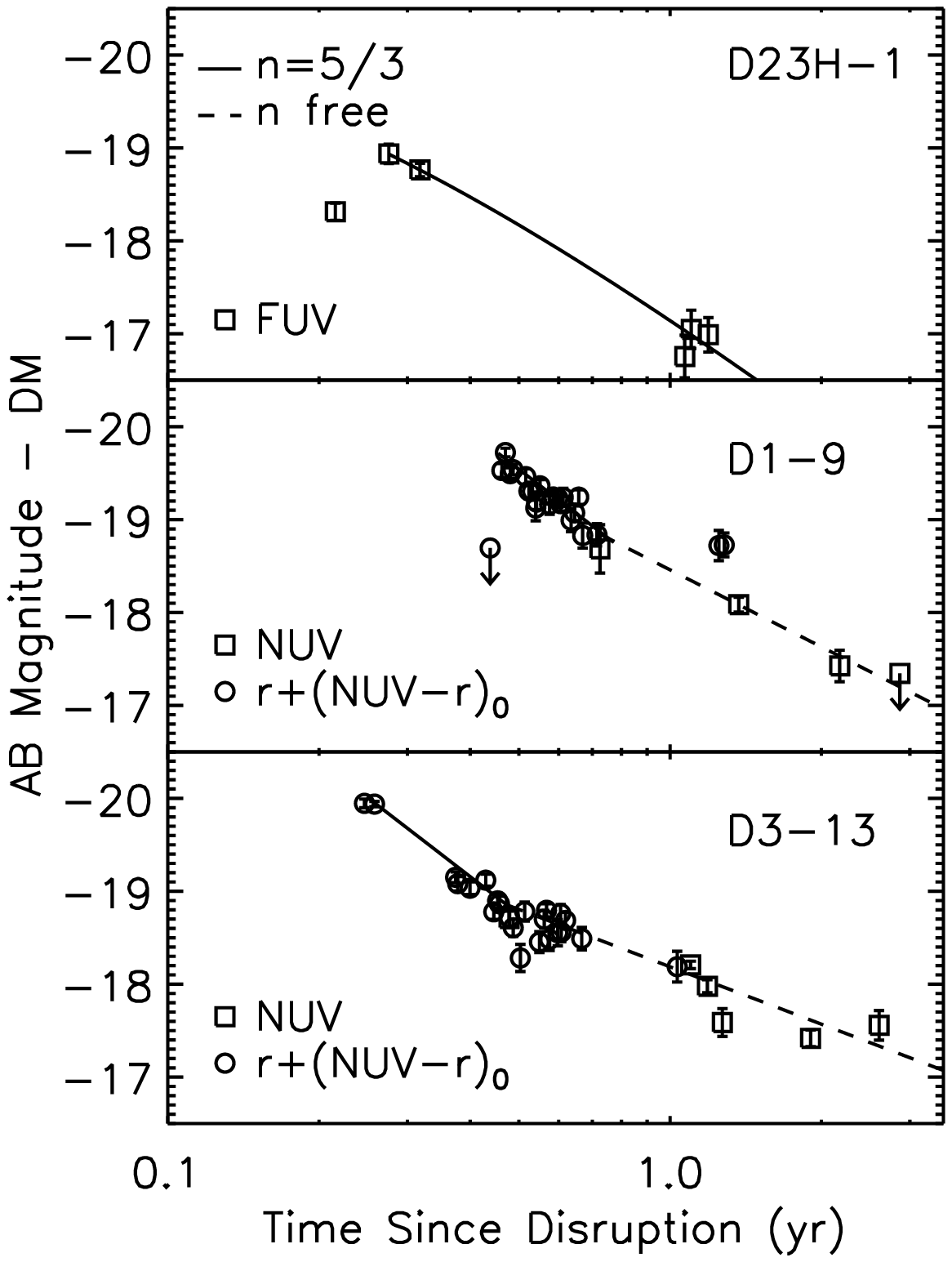}{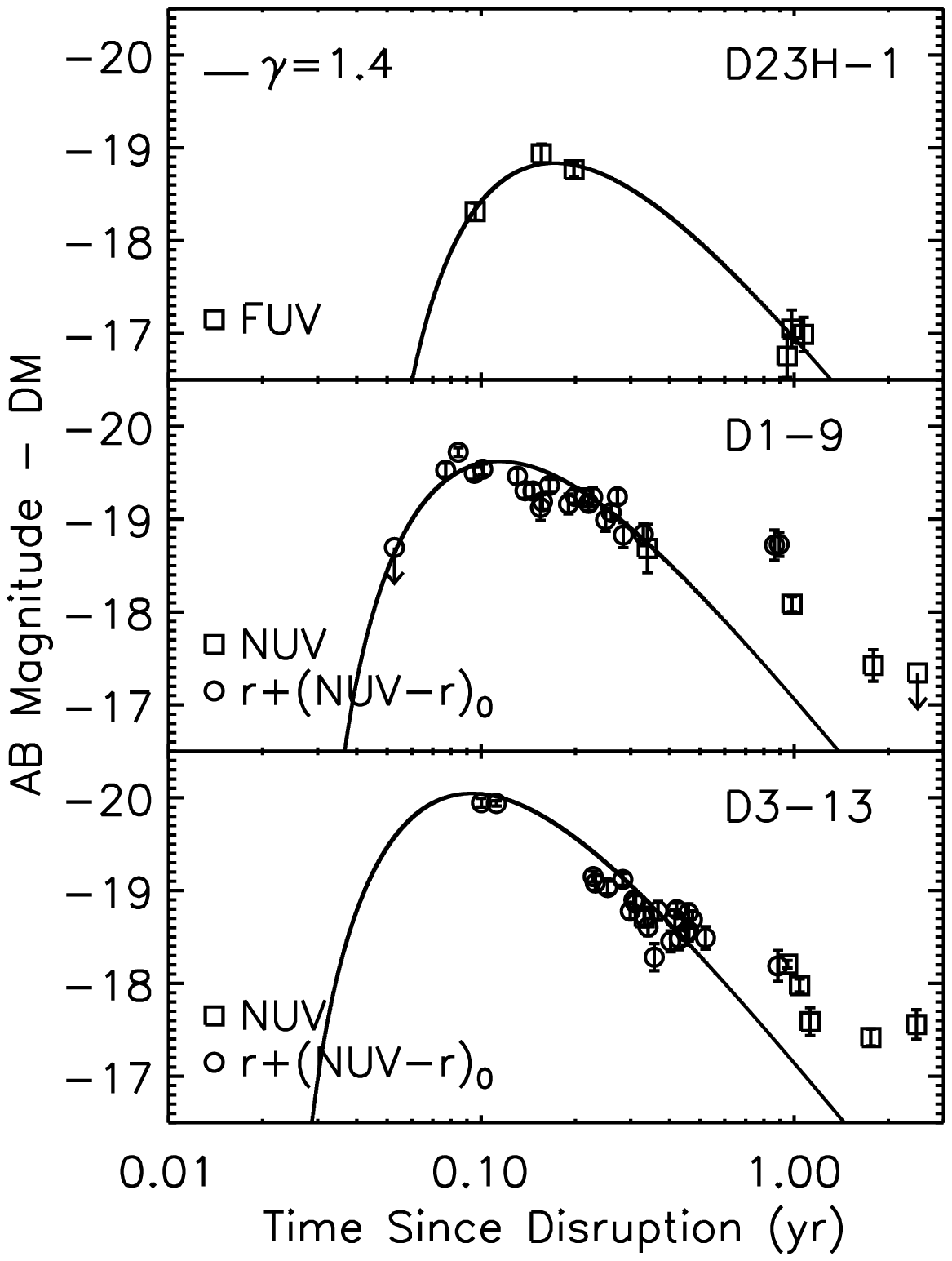}
\caption{
{\it Left:} Light curves of the three \textsl{GALEX} tidal disruption flare candidates, presented in this Paper and Gezari \etal (2008).  The AB magnitudes have been corrected for Galactic extinction, and subtracted by the distance modulus (DM=5 log$(d_{L}/10 $pc)).  The time axis is in years since the time of disruption found in the fit, corrected for time dilation, in the rest frame of the galaxy.  The FUV light curve is shown for D23H-1, and the NUV light curves are shown for D1-9 and D3-13.  The $r$-band light curves are shifted up by the observed average (NUV-$r$) color, (NUV-$r)_{D1-9}=-2.0$ and (NUV-$r)_{D3-13}=-1.4$. The solid lines show the power-law fit with the power-law index, $n$, fixed to 5/3, and the dashed lines show the fits with the power-law index $n$ free to vary.   {\it Right:} the solid line shows the analytical solution from Lodato \etal (2009) of $\dot{M}(t)$ for a solar-type star with a $\gamma=1.4$ polytropic index, scaled in time by $k^{-3/2}(\mbh/10^{6}\msun)^{1/2}$, where we assume $k=3$, and fit for $\mbh = 5.4\times10^{7} , 2.4\times10^{7},$ and $1.2\times10^{7} \msun$ for D23H-1, D1-9, and D3-13, respectively.
 \label{fig:lc}
}
\end{figure*}

The value for the power-law index fitted to the light curve of D23H-1, $n=1.3 \pm 0.9$, has enough flexibility in its error bars to be within $1\sigma$ of $n=5/3$.  The early light curves of the \textsl{GALEX} candidates D1-9 and D3-13, detected in the $g$, $r$, and $i$ bands by CFHTLS in the first $4-5$ months after the peak, were also well fitted with a $n=5/3$ power law, however their late-time light curves traced by the UV photometry showed a break in the power-law to a shallower decay with $n_{D1-9}=1.1 \pm 0.3$ and $n_{D3-13}=0.82 \pm 0.03$ (Gezari \etal 2008).   This behavior at late-times is actually a prediction of tidal disruption models, which predict that after the debris shocks and circularizes, the mass accretion proceeds through viscous processes which should have a characteristic power law of $n=1.2$ (Cannizzo \etal 1990).  The power law fits to the light curves of all three \textsl{GALEX} tidal disruption event candidates as well as fits to the analytically derived L09 curve for a solar-type star are shown in Figure \ref{fig:lc}.  Assuming $k=3$, the fits to the L09 curve have $\mbh = 2.4\times10^{7} \msun$ and $\mbh = 1.2\times10^{7} \msun$ for D1-9 and D3-13, respectively.  To more clearly show the shape of the UV/optical light curves of D1-9 and D3-13, the $r$-band data points are shown along with the NUV data, shifted up by their average NUV$-r$ color: (NUV$-r$)$_{D1-9}=-2.0$ and (NUV$-r$)$_{D3-13}=-1.4$.  The tidal disruption event candidates discovered in the X-rays by \textsl{ROSAT} and \textsl{XMM} Slew Survey (XMMSL1) have sparsely sampled X-ray light curves, but their dramatic fading $3-10$ years after the peak of the flares are also consistent with an $n=5/3$ power-law decay (Komossa \etal 2004; Halpern \etal 2004; Li \etal 2002; Esquej \etal 2008).  

It is surprising how well the light curves of the \textsl{GALEX} candidates are described by a simple power-law decay, and that they are consistent with a power-law index of $n=5/3$ at early times.  However, the value of the power-law index is very sensitive to the time of disruption, and vice versa.  Thus if one wants to use the light curves to measure $t_{0}-t_{D}$ and get an estimate the central black hole mass, or if one wants to measure the power-law index, $n$, to understand the declining mass accretion rate as a function of time, it is more constraining to use models that remove some degrees of freedom in the fit by predicting the rise of the light curve as well as its decay, such as those calculated by L09.  

\begin{figure*}
\plottwo{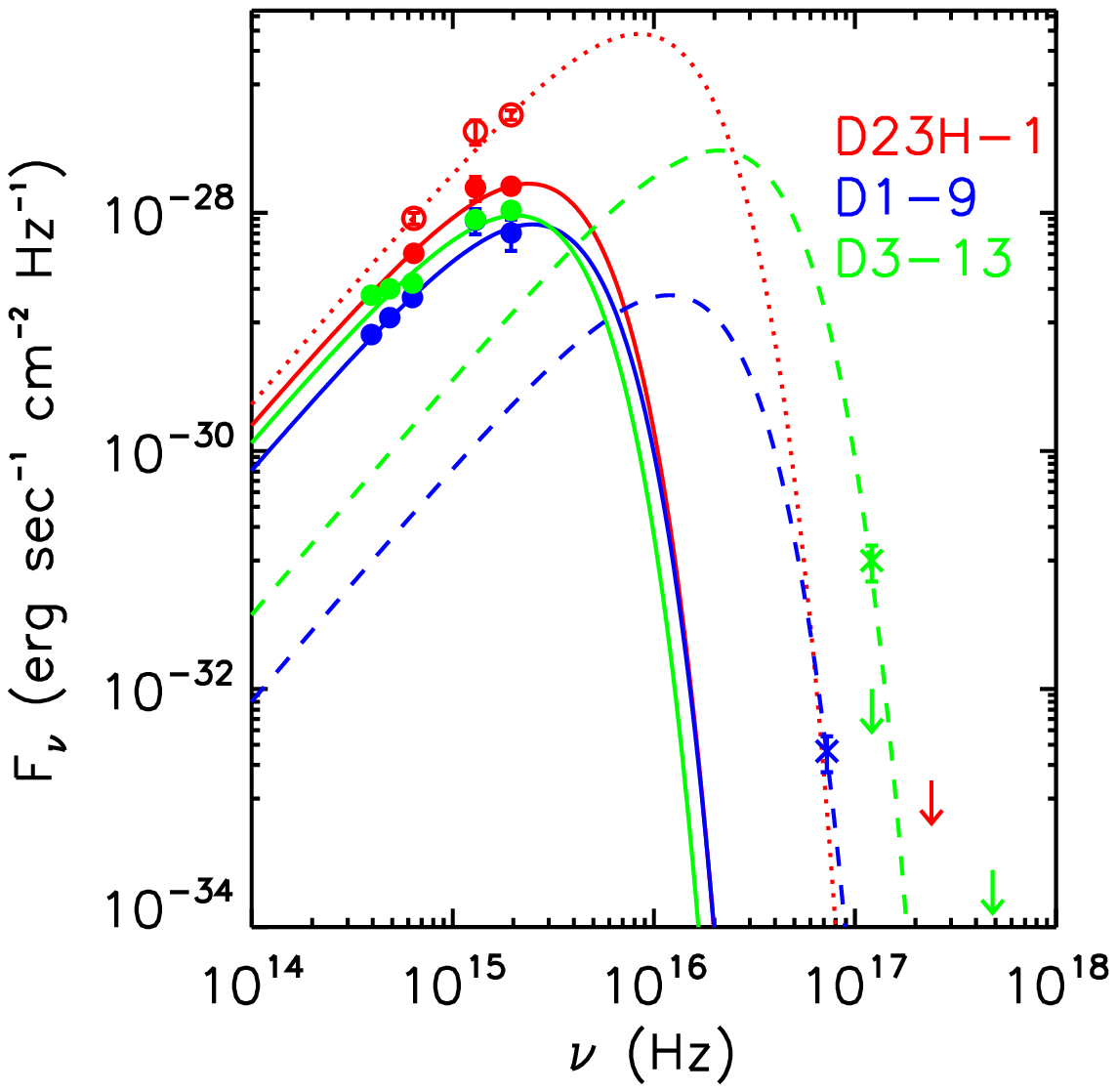}{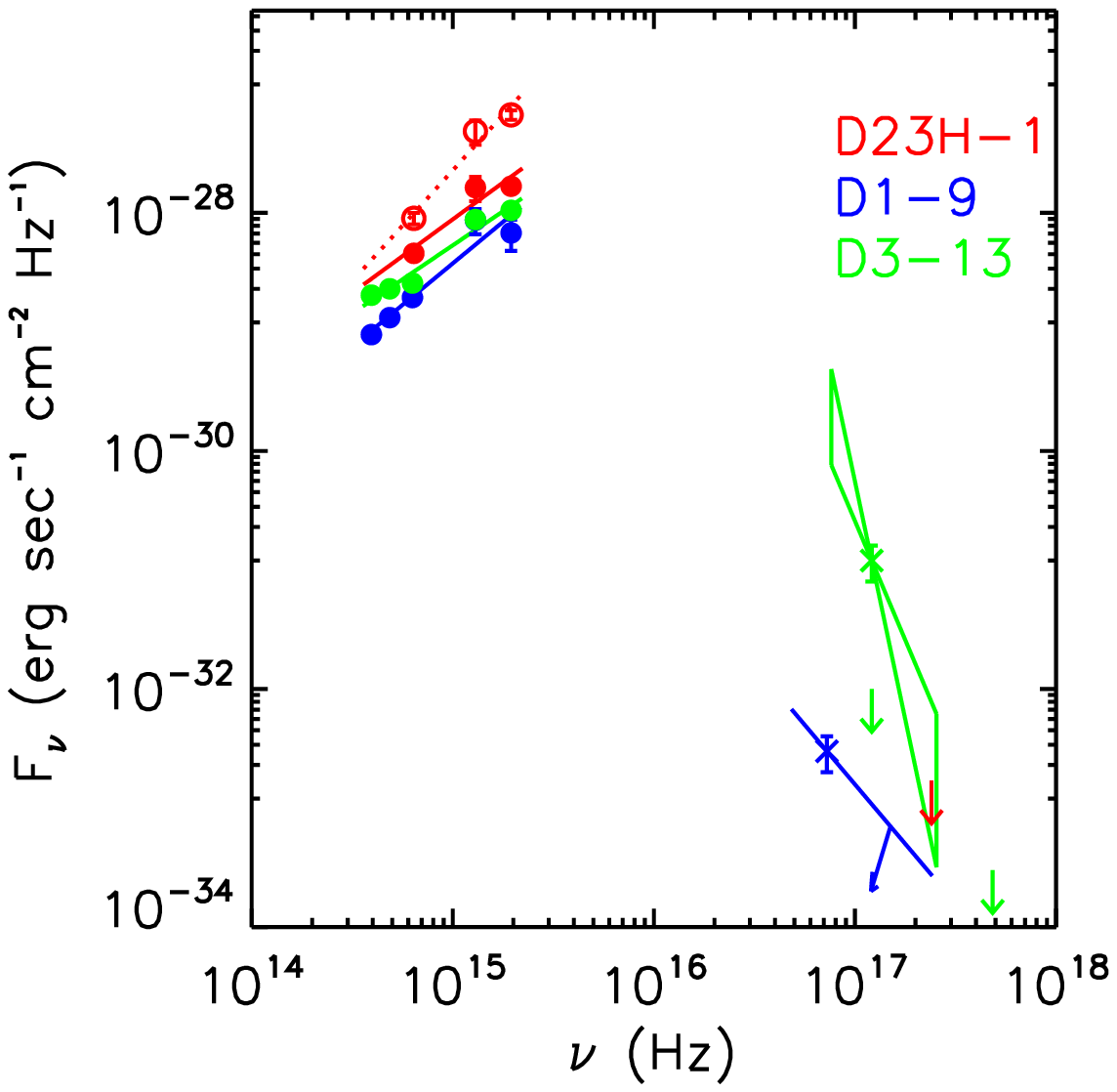}
\caption{
{\it Left}: single temperature blackbody fits to the three \textsl{GALEX} flares.  Fits to the soft X-ray flux densities in D1-9 and D3-13 are shown with a dashed line, and were detected 2.3 and 1.2 yr after the peak of their flares, respectively.  The fluxes corrected for an internal extinction of $E(B-V)_{gas}=0.3$ using the Calzetti (2001) extinction law are plotted with open symbols, and the corresponding blackbody fit is plotted with a dotted line.  {\it Right}: power-law fits to the three \textsl{GALEX} flares and the soft X-ray detections in D1-9 and D3-13.   The fluxes corrected for an internal extinction are plotted with open symbols, and the corresponding power-law fit is plotted with a dotted line.
 \label{fig:sed_bb_pl}
}
\end{figure*}

Shown in Figure \ref{fig:sed_bb_pl} are the blackbody fits and the power-law fits to the UV/optical SEDs of the three \textsl{GALEX} candidates, and in Table \ref{tab:lum} we list the fitted parameters, the reduced $\chi^{2}$ statistic ($\chi_{\nu}$), and the resulting luminosities.  Because the peak of the flares from D1-9 and D3-13 were detected in the optical, with the first UV detection $\sim$four months after the peak, we estimate the simultaneous UV/optical SED by extrapolating the UV fluxes to the time of the peak using the power-law fit to the light curves (shown in Figure \ref{fig:lc}).
We also show the single blackbody fits to the soft X-ray detections for D1-9 and D3-13, that were detected 2.3 and 1.2 yr after the peak, respectively, with $T_{bb}=2.7 \times 10^{5}$ K and $T_{bb}=4.9 \times 10^{5}$ K, respectively, which yield blackbody luminosities of $5.7 \times 10^{43}$ erg s$^{-1}$ and $2.3 \times 10^{45}$ erg s$^{-1}$.  It is important to note that these X-ray detections did not have enough counts to measure the spectrum or its associated blackbody temperature directly, however, the lack of counts $>$ 1 keV in the sources do place constraints on a power-law spectrum, $f_{E} \propto E^{-\Gamma}$, to have $\Gamma > 3$ and $\Gamma = 7\pm2$, respectively, shown in Figure \ref{fig:sed_bb_pl}.

The power-law fits to the UV/optical SEDs of all three \textsl{GALEX} candidates are steep ($\alpha = +1.4$ to $+1.8$), and not typical of an AGN power-law spectrum with $-2.0 < \alpha < -0.43$ (Zheng \etal 1997; Dietrich \etal 2002).  A single blackbody model appears to provide good fit to the UV/optical part of the flare SED of D1-9 ($\chi^{2}_{\nu}=0.97$) and for the internal extinction corrected flare SED of D23H-1 ($\chi^{2}_{\nu}=0.98$).   The poor fit to the UV/optical SED of D3-13 with a single temperature blackbody would be greatly improved with the addition of a second lower temperature blackbody to account for the flattening of the optical SED slope.  However, the blackbody model fits to the UV/optical SEDs need not be unique.  One could imagine that the UV emission is the result of inverse Compton scattering from a hot plasma (corona) with seed photons from synchrotron radiation (perhaps in the radio or infrared).  Future radio and infrared observations would be critical for testing this alternative scenario.  

Now that the census of tidal disruption candidates has reached a total of 10, we can start to compare their properties and investigate the selection effects that arise from their different modes of discovery. 
The RASS detected large amplitude, luminous soft X-ray outbursts from four galaxies with no previous evidence of AGN activity.  The soft X-ray flares were well described by blackbody spectra with $T_{bb}=(6-12) \times 10^{5}$ K, although they could also be fitted with extremely soft power laws with $\Gamma = 3-5$ (Bade \etal 1996; Komossa \& Bade 1999; Komossa \& Greiner 1999; Grupe \etal 2003; Greiner \etal 2000).  More recently a flaring soft X-ray source was reported from a member of the cluster Abell 3571 with a peak luminosity of $L(0.3-2.4$ keV) = $6.8\times10^{42}$ erg s$^{-1}$ and blackbody temperature of $kT_{bb}=0.12$ keV, which decayed by a factor of $>$650 over 10 yr (Cappelluti \etal 2009).  The XMM Slew Survey (XMMSL1) detected large amplitude soft X-ray flares from two galaxies with no previously known AGN.  Only one of the flares had enough counts to be directly fitted with a blackbody with $T_{bb}=1.1\times10^{6}$ K or a power law with $\Gamma = 3$, while the other had counts only in the $0.2-2$ keV band, and was assumed to have $T_{bb}= 8 \times 10^{5}$ K (Esquej \etal 2008).
 
All of the candidates mentioned above have quiescent galaxy hosts, with the exception of NGC 3599 and NGC 5905.  NGC 5905 is a ROSAT candidate that is a starburst galaxy that was found with narrow-slit STIS spectroscopy to host a low-luminosity Seyfert 2 nucleus with $L_{H\alpha} \sim 10^{38}$ erg s$^{-1}$ (Gezari \etal 2003).  NGC 3599 is an XMMSL1 candidate for which follow-up spectroscopy revealed narrow emission lines with ratios on the borderline between a LINER and a Seyfert galaxy (Esquej \etal 2008).  The presence of a low-luminosity AGN in these galaxies does not contradict the tidal disruption scenario, since their amplitudes of $\apgt 100$ in the soft X-rays are more extreme than any previously known cases of AGN variability.

Similarly to NGC 5905, D23H-1 could also harbor a Seyfert 2 nucleus that is masked by the surrounding HII regions in its spectrum.  However, the hard X-ray luminosity upper limits during all three \textsl{GALEX} flares place an upper limit on the presence of a persistent power-law AGN down to $L(2-10)$ keV $< 10^{41}$ erg s$^{-1}$ in D3-13 and D23H-1, and down to $L(2-10$keV)$ < 6 \times 10^{41}$ erg s$^{-1}$ in D1-9.  Although these X-ray luminosity upper limits do not exclude the presence of a low-luminosity AGN in the nuclei of the galaxies, the unique SEDs and light curves of the flares make them stand out from known low-luminosity AGN behavior.

\begin{figure*}
\plotone{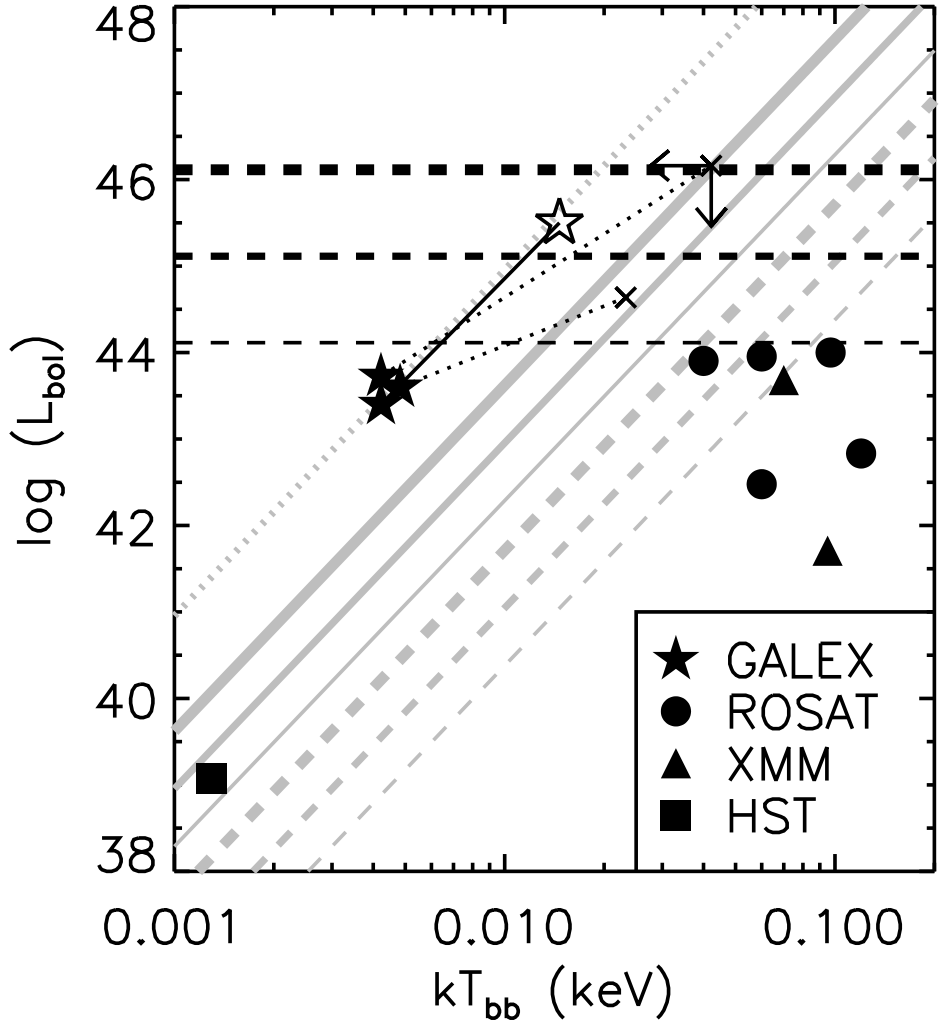}
\caption{
Peak bolometric luminosity of candidate tidal disruption events as a function of their blackbody temperature.  \textsl{GALEX} candidates with soft X-ray detections during their flares have the estimated peak soft X-ray blackbody component plotted with an X, and connected to the UV/optical blackbody component with a dotted line.  The internal extinction corrected blackbody component of D23H-1 is plotted with an open symbol, and connected to the uncorrect blackbody component with a solid line.  The solid gray lines show the luminosity as a function of temperature for a blackbody radius equal to the tidal disruption radius of a $10^6$, $10^7$, and $10^8 \msun$ black hole, and the dotted gray line shows the luminosity for blackbody radius equal to 10 times the radius of a $10^{7} \msun$ black hole.  The dashed gray lines show the same relation, but for $T_{bb}/T_{eff}$ = 3.  The dashed black lines show the Eddington luminosity of a $10^6$, $10^{7}$, and $10^{8} \msun$ black hole.
 \label{fig:lbol}
}
\end{figure*}

If we interpret the steep power-law shapes of the UV and soft X-ray candidates as the Rayleigh Jean's and Wien's tail of a thermal blackbody spectrum, respectively, then their blackbody temperatures, in combination with their bolometric luminosities can be used to infer a characteristic radius of the emission.  Figure \ref{fig:lbol} shows the peak bolometric luminosity and blackbody spectrum temperature ($kT_{bb}$) of the four candidates from RASS summarized in Komossa \etal (2002), the Abell cluster candidate from Cappelluti \etal (2009), two candidates from XMMSL1 reported by Esquej \etal (2006, 2008), the candidate presented in this paper, and the two previous candidates discovered by \textsl{GALEX} reported by Gezari \etal (2006, 2008). The blackbody component of D23H-1 is shown both without the internal extinction correction (solid symbol) and with a correction for $E(B-V)_{gas}=0.3$ (open symbol), and connected with a solid line.  For the \textsl{GALEX} candidates that had a soft X-ray detection during their flare, we connect the UV/optical blackbody component to the soft X-ray blackbody component (plotted with an X) with a dotted line. 
Because the X-ray detections for these candidates were $\sim 1-2$ yr after the peaks of the flares, we scale the fluxes using the light curve fits to estimate their X-ray luminosities during the peak.  However, this is highly uncertain, since the X-ray detection for D3-13 showed variability of a factor of 10 on a 1 day timescale, and may not follow the decay of the UV/optical light curve. We plot the soft X-ray component of D3-13 as an upper limit to account for this variability.   

Also shown is the UV flare detected by \textsl{HST}/Faint Object Camera (FOC) in the nucleus of NGC 4552 (Renzini \etal 1995; Cappellari \etal 1999) interpreted as the accretion of a tidally stripped stellar atmosphere by the mini-AGN at its center.  This galaxy was shown to have another episode of variability 10 yr later, when its NUV flux increased by a factor of 1.2 between epochs separated by 2.4 months (Maoz \etal 2005).  This lower-amplitude variability most likely originates from the low-luminosity AGN detected in the nucleus of the galaxy via broad optical emission lines and a hard X-ray point source with $L_{X}\sim4\times10^{39}$ erg s$^{-1}$ (Cappellari \etal 1999; Xu \etal 2005).  Given the rarity of tidal disruption events, the fact that several other nearby LINERs demonstrated UV variability of a similar amplitude as the flare in NGC 4552 reported a decade earlier by Renzini \etal (1995) casts doubt on its interpretation as a tidal stripping event. 

Shown in comparison are the luminosities expected for a blackbody radius equal to the tidal disruption radius of the central supermassive black hole, using $L_{\rm bol}=4 \pi R_{bb}^{2} \sigma (T_{bb}/f_{c})^{4}$, where $f_{c}\equiv T_{bb}/T_{eff}$ takes into account deviations of the color temperature from the effective temperature due to electron scattering and comptonization in systems with high-mass accretion rates (see Li \etal 2002).  Also shown is the Eddington luminosity of black holes of $10^{6}$, $10^{7}$, and $10^{8} \msun$, where $L_{\rm Edd}=1.3\times10^{44} M_{6}$ erg s$^{-1}$.
There is a large range in color temperatures of the events, with a strong bias toward the wavelength range of the survey used to discover them.  The temperatures do not correlate with the luminosity, as would be expected if there were a single characteristic radius from which the emission was produced.  

The black hole masses inferred from the host galaxies of the candidates using locally established scaling relations between $\mbh$ and the host galaxy bulge, yield black hole masses for the XMMLSS1 candidates of a few times $10^{6} \msun$, $\sim 10^{7} \msun$ for the Abell cluster candidate (Cappelluti \etal 2009), and a few times 10$^{7} \msun$ for the \textsl{GALEX} candidates and the \textsl{ROSAT} candidate NGC 5905.  The low luminosity ($\sim 10^{-3} L_{\rm Edd}$) of NGC 5905 was explained by Li \etal (2002) as a result of the disruption of a brown dwarf or giant planet, while they attributed the small inferred radius as emission from a small patch possibly originating from a narrow stream of accreting material, instead of an optically thick circular torus of material at the tidal disruption radius with $T_{eff} \approx (L_{Edd}/4\pi\sigma R_{T}^{2})^{1/4} = 2.5 \times 10^{5} $K$ M_{6}^{1/12}r_{\star}^{-1/2}m_{\star}^{-1/6}$ (Ulmer 1999).  

In contrast, the blackbody radii inferred for the UV/optical component of the \textsl{GALEX} flares are 10 times the tidal disruption radius of a $10^{7} \msun$ black hole, shown with a dotted line in Figure \ref{fig:lbol}.  The presence of material at such large radii is a natural consequence of the disruption of a star, since at least half of the stellar mass is unbound at high velocities during the disruption process.  An extended envelope of debris may form and reprocess the radiation from the inner accretion torus with an effective temperature of $T_{eff} \approx 2.3 \times 10^{4} $K $(\mbh/10^{7} \msun)^{1/4}$ (Loeb \& Ulmer 1997; Ulmer \etal 1998), close to the temperatures fitted to the UV/optical components of the \textsl{GALEX} flares.  

The events, regardless of what wavelength range they were discovered in, share the properties that (1) their peak accretion rates are large enough to produce high luminosities ($10^{42}$ erg s$^{-1}$), (2) their SEDs appear to be steep power-laws consistent with being thermal in nature, and (3) the power-law decay of their light curves is consistent with the declining mass accretion rate expected from the fallback of debris from a tidally disrupted solar-type star.  However, the geometry of their radiating regions is diverse, and there appear to be two components to their continuum emission.  The soft X-ray components have radii on the order or smaller than the tidal disruption radius of the central black hole, and are most likely associated with the inner accretion torus (or a small region of it).  The rapid variability of the soft X-ray source detected in D3-13 (Gezari \etal 2006, 2008) is consistent with this picture.  While the UV/optical components of the flares appear to have a characteristic radius of roughly $10R_{T}$, possibly tracing an envelope of debris surrounding the black hole.

\section{Conclusions} \label{conc}
We present the third candidate tidal disruption event discovered by \textsl{GALEX}: a UV/optical flare from a star-forming galaxy at $z=0.1855$ with a broadband SED, light curve, and peak luminosity in excellent agreement with the expected properties of a flare from the tidal disruption of a star.  While we can only place upper limits on the presence of a steady-state low-luminosity AGN in the host galaxy down to $L_{X}< 10^{41}$ erg s$^{-1}$, the broadband properties of its flare deviate from the average properties observed for AGNs.  While it is possible that the flares we are detecting with \textsl{GALEX} represent a 
large amplitude, high-luminosity tail of the distribution of UV variability 
from AGNs, such flares would have to comprise of UV emission that is 
not correlated with hard X-ray emission or optical
emission lines (unlike what is observed in the small
number of AGNs that do have detailed optical and X-ray monitoring
campaigns).  Future coordinated optical, UV, and X-ray monitoring of AGNs over wide fields of view will be necessary to determine how favorable this interpretation is over the tidal disruption event scenario. 

The next generation of wide-field synoptic surveys will be at optical wavelengths, and their capabilities for detecting tidal disruption events will depend on how these flares manifest themselves in this wavelength range.  We have now successfully detected three candidates for tidal disruption events from quiescent SMBHs in the optical, and in Gezari \etal (2008) we made favorable predictions for the detection rates for Pan-STARRS and LSST using the theoretically derived black hole mass-dependent volume rate of tidal disruption events that successfully reproduced the observed detection rate by \textsl{GALEX} in the UV.  For this calculation, we assumed that flares at their peak radiated at the Eddington luminosity with $T_{bb}=2.5\times10^{5} M_{6}^{1/2}$ K, but that the fraction of events that produced Eddington luminosity flares declined with increasing black hole mass (Ulmer 1999), and we assumed the light curves followed power-law decays with $n=5/3$, $t_{0}-t_{D}=0.11$ yr $k^{-3/2}M_{6}^{1/2}$, and a spin-up parameter for the star of $k=3$.

Here, we will use a simpler approach to estimate the detection rate of tidal disruption events in optical surveys by using the observed range of Galactic extinction-corrected peak absolute $g$-band magnitudes for the \textsl{GALEX} candidates, $M_{g}(D1-9)=-17.66 \pm 0.05$, $M_{g}(D3-13)=-18.88 \pm 0.05$, $M_{g}(D23H-1)=-17.52\pm 0.12$, where $M_{g}$ at peak for D1-9 and D3-13 is estimated from the peak observed $r$-band magnitude and the average observed $g-r$ color during the flares.  We estimate the black hole mass-dependent volume rate as $N(\mbh)=\dot{N}(\mbh)N_{BH}(\mbh)f(\mbh)$, where $\dot{N}(\mbh)$ is the tidal disruption rate as a function of black hole mass found for elliptical galaxies by Wang \& Merritt (2004), $\dot{N}(\mbh) \sim 1.6 \times 10^{-4}$ yr$^{-1} M_{6}^{-0.3}$, $N_{BH}$ is the black hole mass function estimated from the Ferguson \& Sandage (1991) E+S0 luminosity function scaled to black hole mass using the $\mbh-M_{\rm bulge}$ relation from Merritt \& Ferrarese (2001) and the mean galaxy mass-to-light ratio from Magorrian \etal (1998) with a factor of 2 to account for the bulges of spiral galaxies, and $f(\mbh)=-0.909$log$(\mbh/\msun)+7.00$ is the fraction of flares that have a mass accretion rate close to the Eddington limit for a Salpeter mass function from Ulmer (1999).  The declining fraction of Eddington luminosity flares from higher mass black holes is a result of the decrease in peak accretion rate with increasing black hole mass as $\mbh^{-1/2}$.  If we integrate $\dot{N}(\mbh)$ over the full range of black hole masses, $10^{6}-10^{8}\msun$, this yields $N=1.3 \times 10^{-5}$ yr$^{-1}$ Mpc$^{-3}$.  

For this volume rate, the 50 deg$^{2}$ Pan-STARRS1 Medium Deep Survey (PS1 MDS) will detect 32 events yr$^{-1}$ with $g<22$ out to 790 Mpc (or $z=0.16$).  We have chosen this brightness cut off so that the peak of the flares is at least 2 mag brighter than the detection limit of the survey to allow for the decay of the light curve of the flare to be followed in detail.  This detection rate is double our previous estimate of 15 events yr$^{-1}$.  However, here we have assumed that all flares will have a peak with $M_{g}<-17.5$.  If instead we take into account that these luminosities apply only to the higher mass black holes, i.e., $\mbh > 10^{7} \msun$, then the volume rate for this black hole mass range, $2.3\times10^{-6}$ yr$^{-1}$ Mpc$^{-3}$, corresponds to a detection rate for PS1 MDS of 5 yr$^{-1}$.  This can be thought of as a lower limit, since the survey will be sensitive to fainter events from lower mass black holes at smaller distances. The future LSST will act as a tidal disruption event factory, since its depth and cadence will be similar to PS1 MDS, but with a factor of 400 increase in area that will increase the detection rates by 2 orders of magnitude!

Shallower, wide-field optical time domain surveys such as the PS1 3$\pi$ survey, the Palomar Transient Factory (Rahmer \etal 2008) and Skymapper (Keller \etal 2008) have the potential to detect bright events in the local universe, with detection rates per year of $\approx 1$ event (1000 deg$^{2})^{-1}$ with $g<19$ out to 200 Mpc, or $\approx 1$ event (1000 deg$^{2})^{-1}$ with $r<19$, assuming $g-r>-0.4$, as observed for the \textsl{GALEX} candidates D1-9 and D3-13, $(g-r)_{D1-9}=-0.1$ and $(g-r)_{D3-13}=-0.4$. 

In the near future, we will have the opportunity to exploit the sensitivity, large volume, and daily cadence of the upcoming optical synoptic surveys to accumulate a large number of detailed light curves of tidal disruption events in the local and distant universe.  While each individual tidal disruption event provides a cosmic laboratory to the study the detailed accretion physics around supermassive black holes, the ensemble properties and rates of a large number of events will be a powerful tool for testing the predictions of galaxy stellar dynamical models, and probing the mass function of dormant supermassive black holes in normal galaxies.

\acknowledgements

We thank the anonymous referee for their helpful comments.  S.G. was supported by NASA through Hubble Fellowship grant HST-HF-01219.01-A
awarded by the Space Telescope Science Institute, which is operated by the Association of 
Universities for Research in Astronomy, Inc., for NASA, under contract NAS 5-26555, and in part by \textsl{Chandra} grant G07- 8112X.
We gratefully acknowledge NASA's support for construction,
operation, and science analysis for the \textsl{GALEX} mission,
developed in cooperation with Centre National d'Etudes Spatiales of
France and the Korean Ministry of Science and Technology.
Some of the data presented were obtained at the W.~M.~Keck Observatory, which is operated as a scientific partnership among the California Institute of Technology, the University of California, and NASA.  The Observatory was make possible by the generous financial support of the W.~M.~Keck Foundation.
The analysis pipeline used to reduce the DEIMOS data was developed at 
UC Berkeley with support from NSF grant AST-0071048.
The Hobby-Eberly Telescope (HET) is a joint project of the University
of Texas at Austin, the Pennsylvania State University, Stanford
University, Ludwig-Maximillians-Universit\"at M\"unchen, and
Georg-August-Universit\"at G\"ottingen. The HET is named in honor of
its principal benefactors, William P. Hobby and Robert E. Eberly.
The Marcario Low-Resolution Spectrograph is named for Mike Marcario of
High Lonesome Optics, who fabricated several optics for the instrument
but died before its completion; it is a joint project of the
Hobby-Eberly Telescope partnership and the Instituto de
Astronom\'{\i}a de la Universidad Nacional Aut\'onoma de M\'exico.

\clearpage

\begin{deluxetable}{lcrr}
\tablewidth{0pt}
\tablecaption{\textsl{GALEX} Observations \label{tab:uv}}
\tablehead{
\colhead{UT Date} & \colhead{t$_{exp}$ (ks)} & \colhead{FUV} & \colhead{NUV}
}
\startdata
2004.798 & 17.3 & 22.81 $\pm$ 0.13 & 22.04 $\pm$ 0.20 \\
2005.753 &  6.3 & 22.93 $\pm$ 0.21 & 21.98 $\pm$ 0.20 \\
2006.693 & 12.3 & 23.10 $\pm$ 0.16 & 22.30 $\pm$ 0.20 \\
2007.673 & 16.1 & 21.24 $\pm$ 0.07 & 21.18 $\pm$ 0.19 \\
2007.744 &  6.1 & 20.71 $\pm$ 0.09 & 20.61 $\pm$ 0.19 \\
2007.794 &  7.4 & 20.86 $\pm$ 0.06 & 20.79 $\pm$ 0.20 \\
2008.685 & 10.3 & 22.30 $\pm$ 0.09 & 21.86 $\pm$ 0.20 \\
2008.724 & 10.6 & 22.14 $\pm$ 0.10 & 21.82 $\pm$ 0.20 \\
2008.830 & 17.6 & 22.18 $\pm$ 0.08 & 21.82 $\pm$ 0.20
\enddata
\end{deluxetable}

\begin{center}
\begin{deluxetable}{cccccccccccc}
\footnotesize
\tabletypesize{\scriptsize}
\tablewidth{0pt}
\tablecaption{UV/Optical SED Fits \label{tab:lum}}
\tablehead{
\colhead{ID} & \colhead{$E(B-V)_{int}$} & \colhead{$T_{bb}$} & \colhead{$T_{bb}-1\sigma$} & \colhead{$T_{bb}+1\sigma$} & \colhead{$\chi^{2}_{\nu}(bb)$} & \colhead{$\alpha$} & \colhead{$\chi^{2}_{\nu}(pl)$} & \colhead{log($L_{bb}$)} & \colhead{log($L_{bb}-1\sigma$)} & \colhead{log($L_{bb}+1\sigma$)} & \colhead{log($L_{pl}$)} \\
\colhead{} & \colhead{} & \colhead{($\times 10^{4}$ K)} & \colhead{($\times 10^{4}$ K)} & \colhead{($\times 10^{4}$ K)}  & \colhead{} & \colhead{} & \colhead{} & \colhead{(ergs s$^{-1}$)} & \colhead{(ergs s$^{-1}$)} & \colhead{(ergs s$^{-1}$)} & \colhead{($1350-8500$\AA)} \\
}
\startdata
D1-9 & 0.0 & 5.6 & 4.4 & 6.1 & 0.97 & 1.4 $\pm$ 0.3 & 1.77 & 43.6 & 43.5 & 43.7  & 43.3 \\
D3-13 & 0.0 & 4.9 & 4.7 & 5.1 & 22.48 & 1.1 $\pm$ 0.1 & 14.50 & 43.8 & 43.7 & 43.9  & 43.5 \\
D23H-1 & 0.0 & 4.9 & 4.4 & 5.6 & 0.64 & 1.2 $\pm$ 0.4 & 3.47 & 43.4 & 43.3 & 43.6  & 43.2 \\
D23H-1 & 0.3 & 17 & 12 & 39 & 0.98 & 1.8 $\pm$ 0.4 & 3.12 & 45.5 & 45.0 & 46.8 & 43.7
\enddata
\end{deluxetable}
\end{center}
\end{document}